\documentclass[12pt]{iopart}

\usepackage{iopams}  
\usepackage{graphicx}
\newcommand{\I}[0]{\mathrm{i}}
\newcommand{\V}[1]{\boldsymbol{#1}}
\newcommand{\eqref}[1]{(\ref{#1})}
\newcommand{\be}[0]{\begin{equation}}
\newcommand{\ee}[0]{\end{equation}}
\newcommand{\mean}[1]{\left\langle {#1} \right\rangle}
\newcommand{\paren}[1]{\left( {#1} \right)}
\newcommand{\caja}[1]{\left[  {#1} \right]}
\newcommand{\lazo}[1]{\left\{  {#1} \right\}}
\newcommand{\ket}[1]{\left| {#1} \right\rangle }
\newcommand{\bra}[1]{\left\langle {#1} \right| }
\newcommand{\hsiz}[0]{\hat{\sigma}_i^z}
\newcommand{\hsix}[0]{\hat{\sigma}_i^x}
\newcommand{\siz}[0]{\sigma_i^z}
\newcommand{\six}[0]{\sigma_i^x}
\newcommand{\sump}[0]{\sum_{i=1}^N}

\begin{document}

\title[Many-body transverse interactions  in the quantum annealing]{Many-body transverse interactions  in the quantum annealing of the {\em
    p}-spin ferromagnet}
\author{Beatriz Seoane}
\address{Departamento de F\'\i{}sica
  Te\'orica I, Universidad Complutense, 28040 Madrid, Spain.\\ 
Instituto de Biocomputaci\'on y
  F\'{\i}sica de Sistemas Complejos (BIFI), Zaragoza, Spain.}
\ead{beatriz.seoane@fis.ucm.es}
\author{Hidetoshi Nishimori}
\address{Department of Physics, Tokyo Institute of Technology, Oh-okayama, Meguro-ku, Tokyo 152-8551, Japan}

\begin{abstract}
We study the performance of quantum annealing for the simple $p$-body infinite-range ferromagnetic Ising model. In particular, we generalize the transverse antiferromagnetic interactions proposed by Seki and Nishimori as a quantum driver to many-body transverse interactions to understand if the two-body interactions are essential to allow the system to avoid troublesome first-order quantum phase transitions.  We conclude that the general many-body interactions are effective to let the system evolve only through second-order transitions as long as a few minor conditions are satisfied.  It is also discussed whether the overlap of the ground-state wave function of the new driver term with the target ground state is an essential factor for the success.
\end{abstract}

\maketitle

\section{Introduction}
The task of finding the configuration that optimizes a given cost or energy
function ${\cal H}(\lazo{S_i})$ dependent on a large number $N$ of variables
$S_1,\ldots,S_N$ (often subjected to constraints) is a whole research field by
itself, common to many fields in science. Finding the minimum energy or cost
often becomes a hard task when the constraints in the system, or the
interactions between variables, induce frustration because there is no way to find a
minimum configuration that minimizes the problem locally. The frustration
leads to a rugged landscape of many relative minima, and  an exhaustive
search for the absolute minimum is just not feasible for the interesting sizes (the
dimension of the system often grows exponentially with $N$). As examples of
these optimization problems, one can cite the traveling sales
problem~\cite{papadimitriou:98} or the $k$-SAT problem~\cite{garey:79} in
computer science, or finding the ground state of spin glass
in physics~\cite{mezard:87,nishimori:01}.

Complexity in optimization problems is commonly classified as P if an
algorithm is known to solve the problem in a time that grows polynomially
with $N$. On the contrary, if it is not the case, and the time scales faster
with $N$, these problems are labeled NP and considered as hard problems.
Among all the NP problems, there is a subgroup named NP complete so that any
possible NP problem can be reduced to one of them by means of a polynomial
algorithm. Thus, if one algorithm were found that solved polynomically an NP
complete problem, the whole family of problems would also become easy.  The
problems mentioned above belong all to the NP-complete class.\footnote{With
  the exception of the 2-SAT problem and the 2$D$ Ising spin glass~\cite{barahona:82} that can be
  solved polynomically.}

Statistical mechanics, based on physical intuition, has contributed a lot in
the development of new strategies for optimization problems: parallel
tempering or replica exchange~\cite{hukushima:96}, and simulated
annealing~\cite{kirkpatrick:83} are the two popular and widely used examples. In
this last method, fluctuations are introduced in the problem through a
ficticious temperature. This temperature favors the jump over barriers and
thus encourages the system to visit other possible minima. The system is
then simulated at a temperature $T(t)$ that decreases slowly with time until it
is finally switched off at the end of the simulation. We will refer to this
simulated annealing as classical annealing (CA) in contrast to the quantum
annealing (QA)~\cite{kadowaki:98,finnila:94,das:08,santoro:06}, where fluctuations are induced also
in the system but this time quantum ones. Quantum perturbations allow
tunneling effects, and thus, if narrow enough, barriers can be crossed instead
of surpassed.

In the traditional QA formulation, a time-dependent Hamiltonian is introduced
\be\label{eq:annalgo} \hat H(t)=s(t)\hat H_0+\caja{1-s(t)}\hat V, \ee where $\hat H_0$ is the
target Hamiltonian (or the cost function that one wants to minimize) and
$\hat V$ represents the quantum perturbations. In the field we are working in, the Hamiltonian $\hat H_0$ represents the magnetic interaction between
spins. For the sake of simplicity, we will consider that $\hat H_0$ only
depends on the $z$ components of the Pauli matrix $\hsiz$, where
$i(=1,\ldots,N)$ labels the index of each spin in the system. As normally, we
are interested in finding the lowest energy spin configuration, i.e. the
ground state. Now we introduce the quantum fluctuations through a spin driver
term $\hat V$. In principle, this term is arbitrary, as long
as it does not commute with $\hat H_0$. In
addition, we impose that $\hat V$ has a single, trivial ground state.
A typical example of a driver Hamiltonian is the transverse-field operator
\be\hat V_\mathrm{TF}\equiv -\sump\hsix,\ee where the $\hsix$ $(i=1,\ldots,N)$ are the
$x$ components of the Pauli matrix. This perturbation is very intuitive, since
it represents nothing but the interaction with a magnetic field along the $x$
direction that induces quantum transitions between the eigenstates of $\hat\siz$, whose
modulus is tuned through the control parameter $s(t)$. Initially, at $t=0$, the control parameter
$s(t)$ starts at $s(0)=0$, with $\hat H(0)=\hat V$, and increases
monotonically with time until it reaches unity at time $\tau$ and $\hat
H(\tau)=\hat H_0$. Let us choose the simplest possible scheme where the control
parameter grows linearly with time, i.e. $s(t)=t/\tau$. 

 The evolution of the system, $\ket{\Phi(t)}$, is determined by the
 Schr\"odinger equation, \be\label{eq:Scheq} \I
 \frac{\mathrm{d}}{\mathrm{d}t}\ket{\Phi(t)}=\hat H(t)\ket{\Phi(t)},\,\,0\le
 t\le\tau. \ee The initial state $\ket{\Phi(0)}$ is the ground state of the
 driver Hamiltonian $\hat V$ and is thus known. If the parameter $s(t)$ is
 changed very slowly ($\tau$ is very long), the state will be at every time
 very close to the instantaneous ground state. If it so, by tuning the
 parameters, one will move adiabatically from the initial ground state to the
 ground state of $\hat H_0$.

 The adiabatic theorem states that the system stays close to the instantaneous
 ground state as long as $\tau\gg \Delta^{-2}_\mathrm{min}$ where
 $\Delta_\mathrm{min}$ is the minimum energy gap from the ground state. Of
 course, in order for the above argument to be of general use, this
 $\Delta_\mathrm{min}$ cannot decrease with $N$ too fast. In fact, if the
 energy gap decays exponentially with the system size, as happens generally in
 first-order transitions, the running time will increase exponentially with
 $N$ and the QA would not help to solve the problem efficiently.

This vanishing exponential gap present in many first-order transitions is
sometimes considered to be one of the most important drawbacks of quantum
annealing. Its presence was somehow shadowed for certain time by the
preasymptotic behavior displayed in the small system sizes feasible in
simulations~\cite{farhi:01,hogg:03,young:08}. Indeed, in the last years, an
increasing number of first-order transitions in the annealing parameters are
being found~\cite{young:10,hen:11,jorg:08,jorg:10a,jorg:10b}. It has thus been
suggested that the presence of these quantum first-order transitions when
tuning the transverse field is an intrinsic property of the systems with
complicate free energy landscape, i.e. the hard problems, leading a
pessimistic scenario for the QA algorithm~\cite{young:10,hen:11,jorg:08,jorg:10a,jorg:10b}.

Recently, it was found that the ferromagnetic $p$-spin model, a model without
disorder and with a simple free energy landscape, also suffers from this kind of
first-order transition~\cite{jorg:10a}.  Due to its simplicity, this model
constitutes a perfect benchmark to study the QA performance. Indeed, it was
recently shown~\cite{seki:12} that, at least for finite values of $p$ and
$p\neq 3$, it is possible to avoid this first-order transition by appending an
additional antiferromagnetic driver term and performing the annealing along a
curve in a space of two annealing parameters instead of just one. This study
changes the paradigm about first-order transitions in QA, since the failure of
QA strategies observed up to now could be a failure of the standard
formulation of QA with a transverse field, not a failure of the algorithm
itself.

Here we go deeper into this problem, studying a family of alternative driver
terms, displaying different symmetries. We show analytically the existence of
paths that cross only a second-order transition and thus the speed of QA is
not exponentially damped.  Indeed, in a second order transition the gap 
  vanishes only polinomially with the number of particles, which must be
  compared with the exponential damping observed in the first order transition. The solution to the problem is not unique
and we study the properties of these new driver terms, reaching the conclusion
that the structure of the ground state of the additional Hamiltonians is not
the main important feature that makes the whole algorithm success as argued
in~\cite{bapst:12}.

\section{Problem}
Our starting point is the ferromagnetic $p$-spin model ($p=2,\,3,\,4\ldots$)
\begin{equation}\label{eq:H0}
\hat{H}_0=-N\paren{\frac{1}{N}\sump\hsiz}^p.
\end{equation}
The ground state for this model, $\ket{\Phi_0}$, corresponds to the state of
all the spins aligned along the $z$ direction. In order to avoid the
degeneracy of the up and down configurations present in even powers of $p$, we
consider here only the odd values of $p$ and $p\ge 3$.  In the limiting
$p\to\infty$ case, this model is nothing but the Grover
problem~\cite{jorg:10a,grover:97}.  Although the Grover's quantum
  algorithm,  whose reformulation in quantum annealing is given
  in~\cite{roland:03}, is considered a success of the quantum
  algorithm (provides a square-root gain with respect to the
  classical search~\cite{grover:96}) it remains being a hard problem even with
quantum algorithms.  Now we consider the problem of finding
this already known ground state $\ket{\Phi_0}$ of \eqref{eq:H0} with the QA
algorithm using two driving terms.

As usual, we consider the traditional transverse field
operator, 
\be\label{eq:VTF}\hat V_\mathrm{TF}\equiv -\sump\hsix,\ee 
whose ground state,
$\ket{\Phi^\mathrm{TF}}$, is the one where all the $N$ spins are pointing to the
positive direction along the $x$ axis. 
We next introduce a second Hamiltonian inspired in the antiferromagnetic
interaction suggested in \cite{seki:12},
\begin{equation}\label{eq:Vk}
\hat{V}_k=+N\paren{\frac{1}{N}\sump\hsix}^k,
\end{equation}
that depends on a parameter $k(>1)$. When $k=2$, we recover the
antiferromagnetic interaction studied in \cite{seki:12}.  The ground state for
this Hamiltonian, namely $\ket{\Phi_k}$, depends on the value of the power
$k$. When $k$ is odd, the energy is minimum when all spins are aligned along
the $x$ axis but pointing to the negative direction.  On the contrary, when
$k$ is even, the ground state corresponds to the state with total
$\sump\six=0$ if $N$ is odd, or $\sump\six=\pm 1$ for $N$ even. One of
the goals of the present paper is to clarify whether the value $k=2$ is
essential to avoid the first-order transition.

If we sum up  \eqref{eq:annalgo}, \eqref{eq:VTF} and \eqref{eq:Vk}, the new Hamiltonian of the problem reads as
\begin{equation}
\label{eq:H}
\hat{H}(s,\lambda)=s\caja{\lambda\hat{H}_0+(1-\lambda)\hat{V}_{k}}+(1-s)\hat{V}_{\mathrm{TF}}.
\end{equation}
Here there are two annealing parameters, $s$ and $\lambda$. These parameters
will be tuned slowly during the annealing process so that, at the final time,
$\tau$, $s(\tau)=\lambda(\tau)=1$ and the target Hamiltonian \eqref{eq:H0} is
thus recovered. In that way, one can explore the annealing process following
infinitely different paths. It might resemble the idea of nondeterministic
Turing machines, but one must always keep in mind that, even though many paths
are possible,  only one is chosen in each particular realization.

The traditional QA is one of the infinite possible paths in \eqref{eq:H}. In
fact, one can remove the influence of $\hat V_k$, just by fixing
$\lambda(t)=1$. Then, the annealing is performed by tuning $s$ from 0 to 1.
If one looks at the configurations, at $t=0$ all spins should be aligned with
the $x$ axis, and at the end, with the $z$ axis. In this case, we know that
the system suffers from a quantum first-order phase transition between
these two states. This transition ruins the efficiency of the algorithm as it
becomes exponential~\cite{jorg:10a}. The idea of introducing this
two-parameter space $(\lambda,s)$ is precise to try avoid this transition by
following an alternative route.  Seki and Nishimori succeeded in finding
ingenious paths~\cite{seki:12} with antiferromagnetic interactions, and here,
we generalize that method to check how the value of $k$ affects the
conclusion.

\section{Analysis by a semi-classical approach}\label{sec:CA}

The QA strategy will succeed if we are able to find a path in the space of
parameters $(\lambda,s)$ that avoids crossing any first-order transition. With
this aim, we compute in this section the phase diagram correspondent to the
new Hamiltonian \eqref{eq:H}, as a function of the parameter $k$. The
$N\to\infty$ limit can be computed analytically using a semi-classical
approximation (method to be explained below) or the Trotter-Suzuki
decomposition formula~\cite{suzuki:76} and the static approximation (see
\ref{sec:SA}), leading to equivalent results.

\subsection{General Properties}

As a starting point, let us rewrite the Hamiltonian \eqref{eq:H} in terms of the total 
spin variables ($S^\alpha=\frac{1}{2}\sump\sigma_i^\alpha$ with $\alpha=x,\,y$ and $z$),
\be
\label{eq:Hnew}
\hat{H}(s,\lambda)=-s\lambda
N\paren{\frac{2}{N}S^z}^p+s\,(1-\lambda)N\paren{\frac{2}{N}S^x}^k-2(1-s)S^x.
\ee This Hamiltonian commutes with the total squared spin, $S^2$. Since
  the total spin is conserved and the initial state in the annealing process
  is the one with all spins aligned with the $x$ axis, we are only interested
in studying the maximum possible $S$ value, i.e. $S=N/2$.

Now, consider the normalized variables 
$m^\alpha=S^\alpha/S$, with $\alpha =x,\,y$ and $z$. The commutation
relations for these variables are
\begin{eqnarray}
[m^x,m^y]=\I\frac{2}{N}m^z,
\end{eqnarray}
and cyclic permutations. The normalized variable $m^\alpha$ can take
$N+1$ values within the interval $[-1,1]$. Thus, in the large $N$ limit, these
variables commute, and we can consider them as the components of a classical
unit vector,
i.e. $\V{m}=(\cos\theta,\sin\theta\sin\varphi,\sin\theta\cos\varphi)$,
being $\theta$ the polar angle measured from the $x$ axis, and
$\varphi$ the azimuthal one measured from the $z$ axis.

Considering the system now as classic, we can write the energy per spin as \be
e=-s\lambda
(\sin\theta\cos\varphi)^p+s\,(1-\lambda)\cos^k\theta-(1-s)\cos\theta.  \ee The
equilibrium state will be determined by the minimum of $e$. Since $p$ is odd,
the minimum lies on the plane with $\varphi=0$, which we call  $XZ^+$ plane.  The energy on this
plane is labeled only by the polar angle $\theta$ \be\label{eq:mine}
e=-s\,\lambda \sin^p\theta+s\,(1-\lambda)\cos^k\theta-(1-s)\cos\theta.  \ee We
search the $\theta_0\in[0,\pi]$  that minimizes \eqref{eq:mine}\footnote{Negative magnetizations along $z$
  axis have always higher free energy due to the change of sign in the
  $\sin^p\theta$ term in \eqref{eq:mine} (remember that we only consider the $p$ odd case in
  this work).}.  The condition for the minimum
is \be\label{eq:condmin} \frac{\partial e}{\partial\theta_0}=-p\,s\,\lambda
\sin^{p-1}\theta_0
\cos\theta_0-k\,s\,(1-\lambda)\cos^{k-1}\theta_0\sin\theta_0+(1-s)\sin\theta_0=0,
\ee whose solutions are the angles $\theta_0$ that satisfy either
$\sin\theta_0=0$ or \be\label{eq:eqtheta0} p\,s\,\lambda \sin^{p-2}\theta_0
\cos\theta_0+k\,s\,(1-\lambda)\cos^{k-1}\theta_0-1+s=0. \ee These two
equations have more than one solution, and each one corresponds to a different
phase. We will consider them as ferromagnetic if $m^z(=\sin\theta_0)>0$, and
quantum paramagnetic if $m^z=0$.  The most stable one at each point
$(\lambda,s)$ will be the absolute minimum of $e$.

We begin with the quantum paramagnetic solutions. The equation
$\sin\theta_0=0$ is satisfied for $\theta_0=0$ or $\pi$. The case $\theta_0=0$
corresponds to positive $x$ magnetization, $m^x=1$. We name this phase QP$^+$.
Its energy is obtained by inserting this angle in \eqref{eq:mine},
\be\label{eq:fQPpca} e_{\mathrm{QP}^+}(s,\lambda)=s(1-\lambda)-1+s.  \ee The
other paramagnetic solution, $\theta_0=\pi$, corresponds to negative
magnetization, $m^x=-1$. We call this phase QP$^-$. This phase is only stable
for odd values of $k$ and its energy is \be \label{eq:fQPmca}
e_{\mathrm{QP}^-}(s,\lambda)=-s(1-\lambda)+1-s. \ee This phase will not appear
in the phase diagrams for $k$ even, since its energy is always positive in
the range of parameters $0\le s,\,\lambda\le 1$.

We consider next the ferromagnetic solutions ($\theta_0> 0$). The purely
ferromagnetic solution $\sin\theta_0=1$ is only a valid solution on the line
$s=1$. Apart from this line, equation \eqref{eq:eqtheta0} cannot be
explicitly solved for any value of $p$, but it can be done in the $p\to\infty$
limit. We study below all the solutions for this limit
and discuss their validity for $p$ finite.

\subsection{Phase diagram for $p\to\infty$}

In this limit, \eqref{eq:eqtheta0} has two possible ferromagnetic solutions.
The parameter $p$ appears in \eqref{eq:eqtheta0} through
$p\sin^{p-2}\theta_0$. We consider the two possible limits for the sine power,
$\sin^{p-2}\theta_0\to$1 (for the F phase) and 0 (for the F' phase), always
keeping $\theta_0> 0$.

We begin the discussion with the F phase. With this aim, we assume
\be\label{eq:limF} \sin^{p}\theta_0\to1,\ee and substitute it in
\eqref{eq:eqtheta0}, \be\label{eq:eqF1}
p\,s\,\lambda\cos\theta_0+k\,s\,(1-\lambda)\cos^{k-1}\theta_0-1+s=0.  \ee In
the $p\to\infty$ limit, this equation can only be satisfied if either the
cosine vanishes, i.e.
$\theta_0=\pi/2$ (but only on the line $s=1$), or $p\cos\theta_0$ tends to a
constant. Let us investigate this second case. We consider $\cos\theta_0=c/p$,
with $c$ a $p$-independent constant, and introduce it in \eqref{eq:eqF1}, and
taking the $p\to\infty$ limit, the equation reads \be\label{eq:eqF2}
s\,\lambda\,c-1+s=0, \ee whose solution is $c=(1-s)/s\lambda$. Thus,
\be\label{eq:solFpinfty} \cos\theta_0=\frac{1-s}{s\,p\,\lambda}\to 0, \ee is a
solution to \eqref{eq:eqtheta0}. Still we need to check that this
$\theta_0$ agrees with the initial assumption \eqref{eq:limF}. Indeed,
\begin{eqnarray}\nonumber
\lim_{p\to\infty} \sin^{p-2}\theta_0=\lim_{p\to\infty}
\caja{1-\paren{\frac{1-s}{2\,s\,p\,\lambda}}^2}^{p}=1.
\end{eqnarray}
We obtain the energy for this phase introducing \eqref{eq:solFpinfty} in \eqref{eq:mine}
\be\label{eq:fFpinftyca}\left. e_\mathrm{F}(s,\lambda)^{k}\right|_{p\to\infty}=-s\,\lambda.\ee

On the other hand, the F' solution is obtained assuming the opposite limit,
\be\label{eq:limF'} p\sin^{p}\theta_0\to0.\ee Under this assumption,
\eqref{eq:eqtheta0} reduces to \be k\,
s\,(1-\lambda)\cos^{k-1}\theta_0-1+s=0, \ee whose solution is
\be\label{eq:solFpCA} \cos\theta_0=\caja{\frac{1-s}{k\,
    s\,(1-\lambda)}}^{\frac{1}{k-1}}. \ee 
Note that if $k$ is odd, the negative solution for the cosine is also a
valid solution. However, it has always  higher energy than its positive
counterpart, so we will not consider it for further discussions.

The energy for the F' phase when $p\to\infty$ is
\be\label{eq:freeFpCAifnty}
\left. e_{\mathrm{F'}}(s,\lambda)\right|_{p\to\infty}=-\frac{k-1}{k}\caja{\frac{1-s}{k\,s\,(1-\lambda)}}^\frac{1}{k-1}(1-s).\ee

\begin{figure}
\begin{center}
 \includegraphics[angle=270,width=0.45\columnwidth,trim=10 20 10
   20]{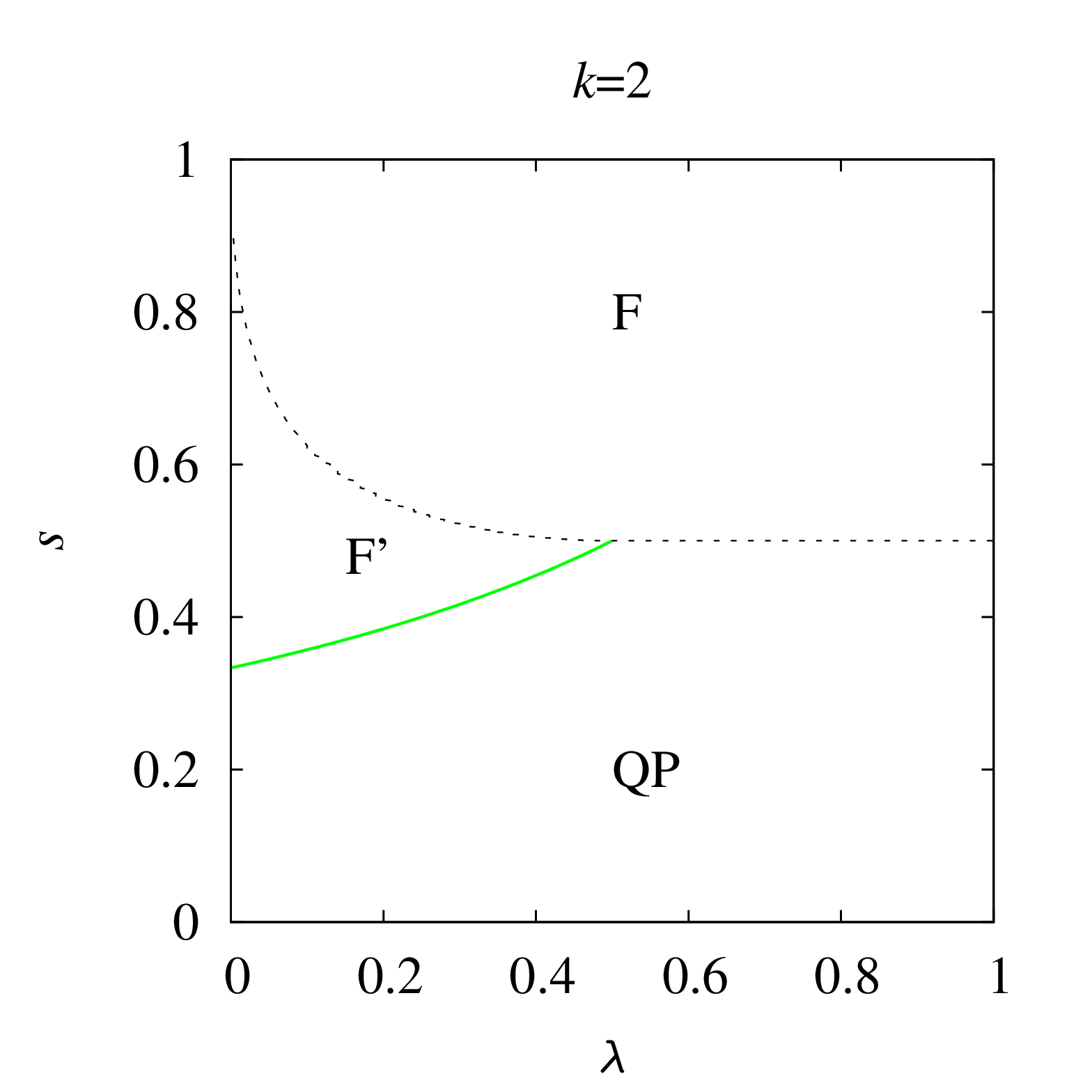}
 \includegraphics[angle=270,width=0.45\columnwidth,trim=10 20 10
   20]{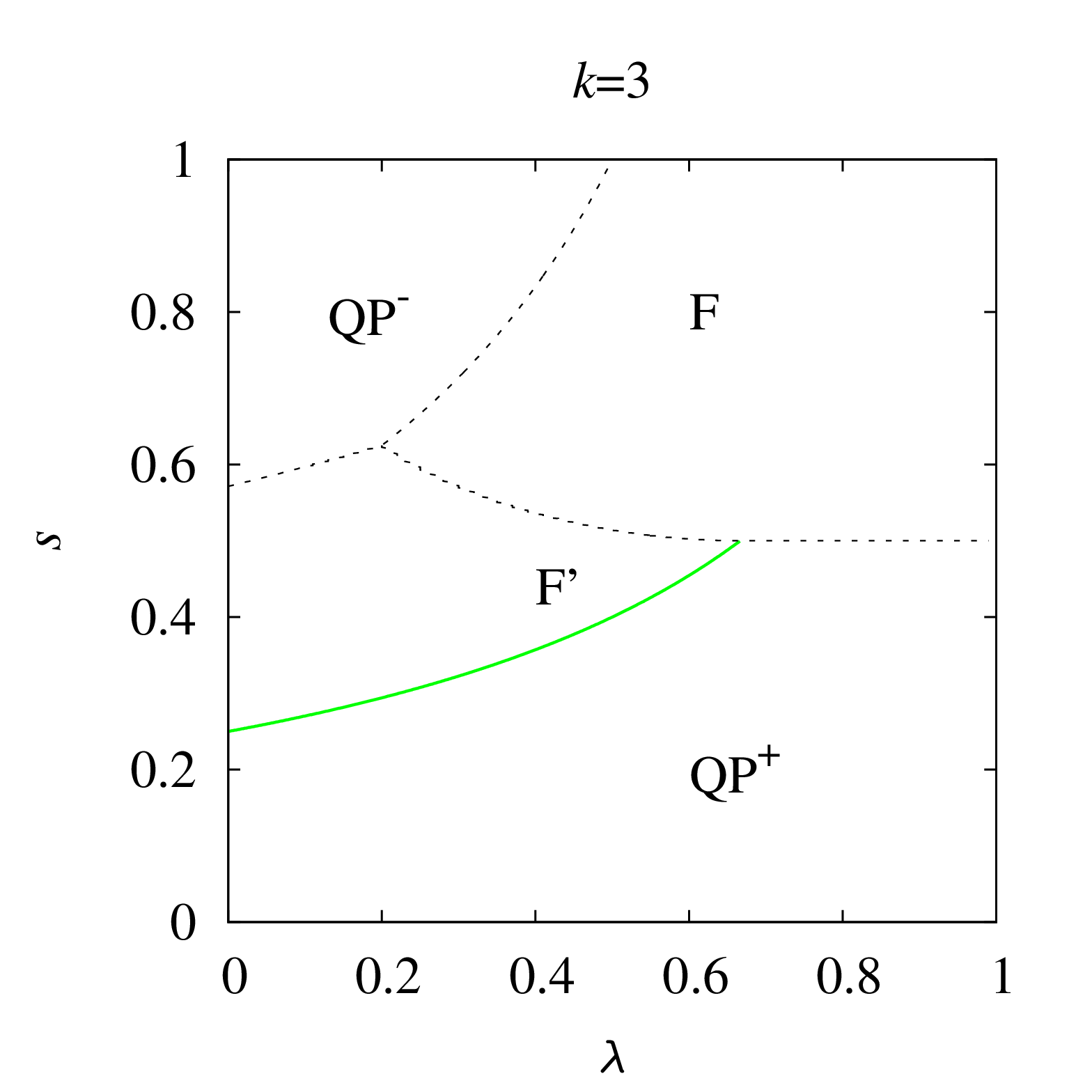}
 \includegraphics[angle=270,width=0.45\columnwidth,trim=10 20 10
   20]{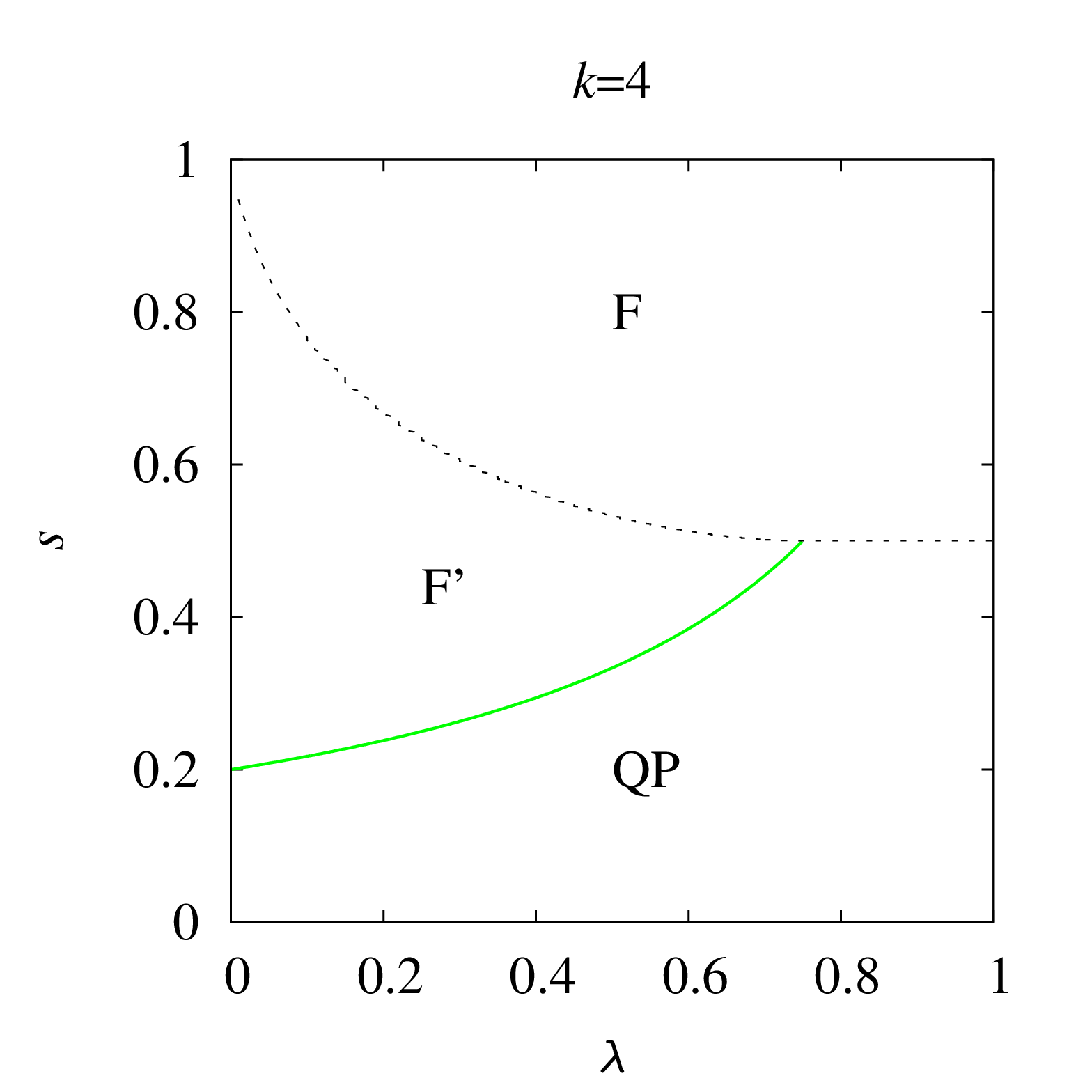}
 \includegraphics[angle=270,width=0.45\columnwidth,trim=10 20 10
   20]{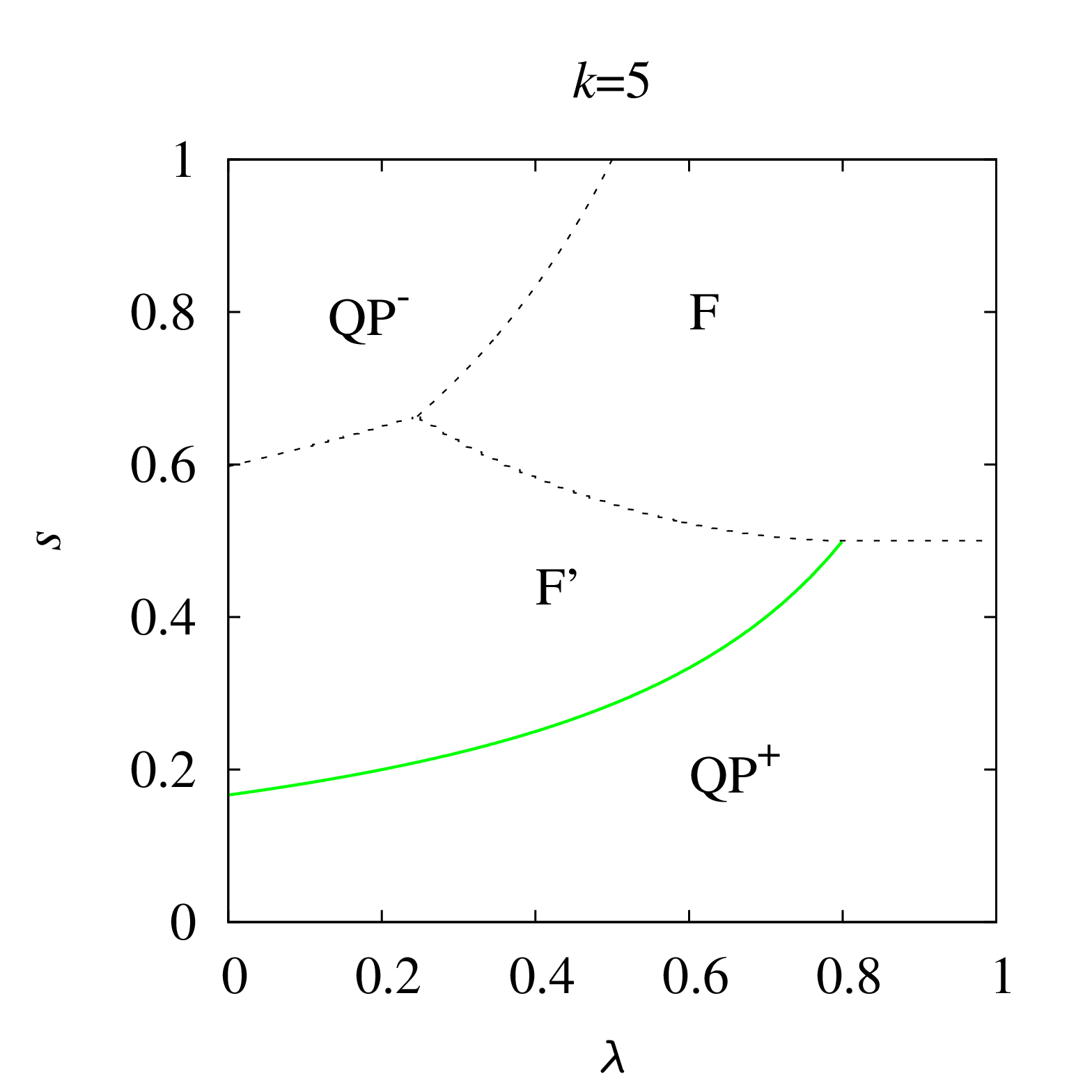}
\caption{Phase diagram for $p\to\infty$. Dashed black lines represent first-order transitions, whereas the solid line in light green accounts for the
  second-order transition.}\label{fig:pd-pifty}
 \end{center}
\end{figure}

Up to this point, we have obtained all the possible solutions to
\eqref{eq:eqtheta0} in the $p\to\infty$ limit: three (for even $k$) and four (for odd $k$) phases. We can use
the energies to determine which phase is the most stable at each point $(\lambda,s)$. We show in figure \ref{fig:pd-pifty} several phase diagrams for $k=2,\,3,\,4$ and $5$.
Let us analyze the nature of each transition. We begin with the transition line
between the F' and  QP$^+$ phases. This line is obtained by solving
$e_\mathrm{F'}-e_\mathrm{QP}=0$ using the expressions \eqref{eq:freeFpCAifnty}
and \eqref{eq:fQPpca}. This equality is fulfilled on the line
$s=1/[1+k(1-\lambda)]$. On this line, $m^x=\cos\theta_0=1$ in both phases,
which corresponds to a second-order transition.  On the other hand, the
transition between the F and the QP$^+$ phases lies on the $s=1/2$ line and,
since magnetization is discontinuous, it is first-order. The
second-order transition extends from $(\lambda,s)=(0,1/(k+1))$ to
$(\lambda,s)=((k-1)/k,1/2)$, the point where these two kinds of transitions
cross. According to that, the higher $k$ is, the broader the second-order line
and the smaller the QP$^+$ region are. Furthermore, in the $k\to\infty$ limit,
the QP$^+$ region completely disappears.

Still there is a first-order transition between the F and  F' phases,
determined by the solution of $e_\mathrm{F'}-e_\mathrm{F}=0$ using
\eqref{eq:freeFpCAifnty} and \eqref{eq:fFpinftyca}. We solve this equation
numerically and obtain the curve displayed in figure  \ref{fig:pd-pifty}.
On this line, the magnetizations are discontinuous but at the point
$(\lambda,s)=(0,1)$ where they two become equal, 
$m^z=\sin\theta_0=1$. The transition is then first-order, but in the mentioned 
point, where it would be second-order.

Up to this point, the discussion is common for even and odd values of
$k$. However, in this latter case the QP$^-$ phase also exists.  Thereby, two
additional transitions between F or F' phases and the QP$^-$ phase appear. In
both cases the $x$ magnetization changes the sign on the transition, and then, they
are first-order. The transition lines are obtained by solving the equations
$e_\mathrm{QP^-}-e_\mathrm{F}=0$, leading to $s=1/(2(1-\lambda))$, and
$e_\mathrm{QP^-}-e_\mathrm{F'}=0$ which must be solved numerically. We display
all the transition lines in figure \ref{fig:pd-pifty}. 

According to these results, when we consider the $p\to\infty$ limit, there is
only one single path that succeeds in avoiding first order transitions. This
is the straight line that joins the initial point $(\lambda,s)=(0,0)$ with the
left upper corner, $(0,1)$, and the final state $(1,1)$. However, even though
this path only crosses second order transitions, along this way there is no
quantum annealing process, as can be seen by an insertion of these parameter
values into the Hamiltonian \eqref{eq:H}, and thus this path is meaningless.

\section{Phase Diagram}
\begin{figure}
\begin{center}
 \includegraphics[angle=270,width=0.6\columnwidth,trim=10 20 10
   20]{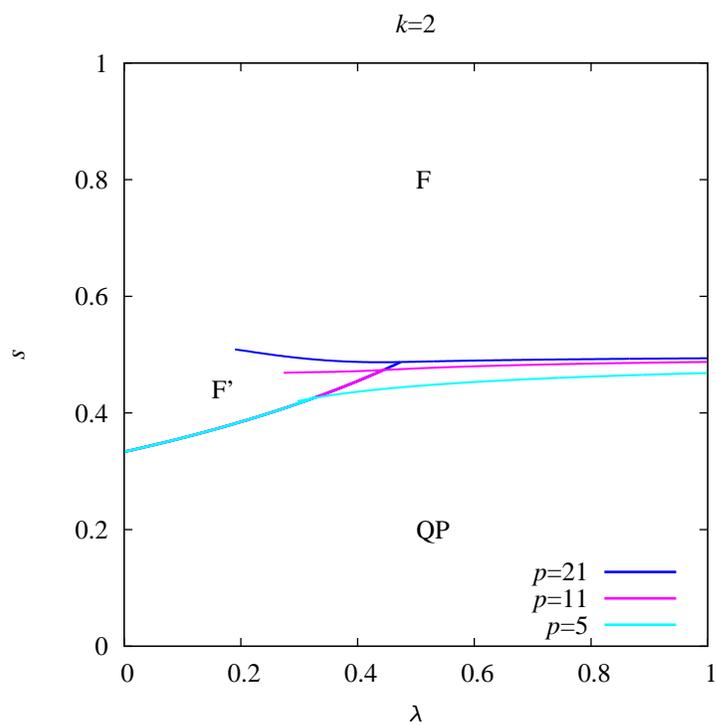}
\caption{Phase diagram for $k=2$. This is the same phase diagram as in reference \cite{seki:12}.  The transition between the F' and QP phases is of second order, and the F-QP and F-F' transitions are of first order. }\label{fig:phase-diag-k2}
 \end{center}
\end{figure}

\begin{figure}
\begin{center}
 \includegraphics[angle=270,width=0.6\columnwidth,trim=10 20 10
   20]{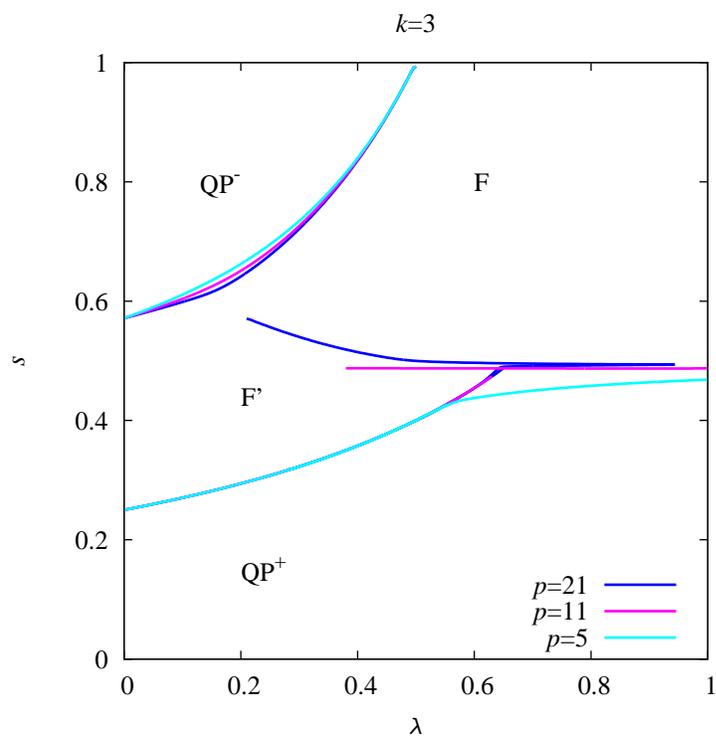}
\caption{Phase diagram for $k=3$. Only the F'-QP$^+$ transition is of second order.}\label{fig:phase-diag-k3}
 \end{center}
\end{figure}

\begin{figure}
\begin{center}
 \includegraphics[angle=270,width=0.6\columnwidth,trim=10 20 10
   20]{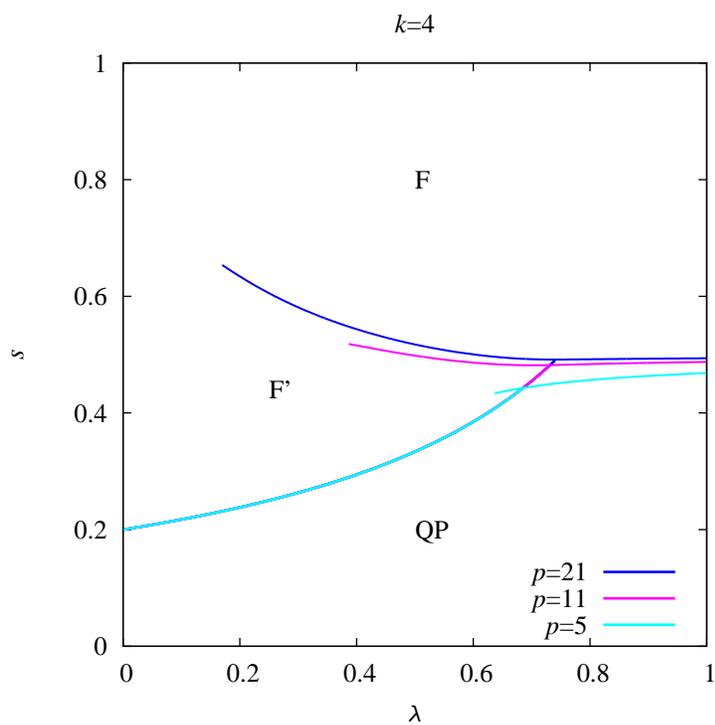}
\caption{Phase diagram for $k=4$. The structure is qualitatively the same as for $k=2$. }\label{fig:phase-diag-k4}
 \end{center}
\end{figure}

\begin{figure}
\begin{center}
 \includegraphics[angle=270,width=0.6\columnwidth,trim=10 20 10
   20]{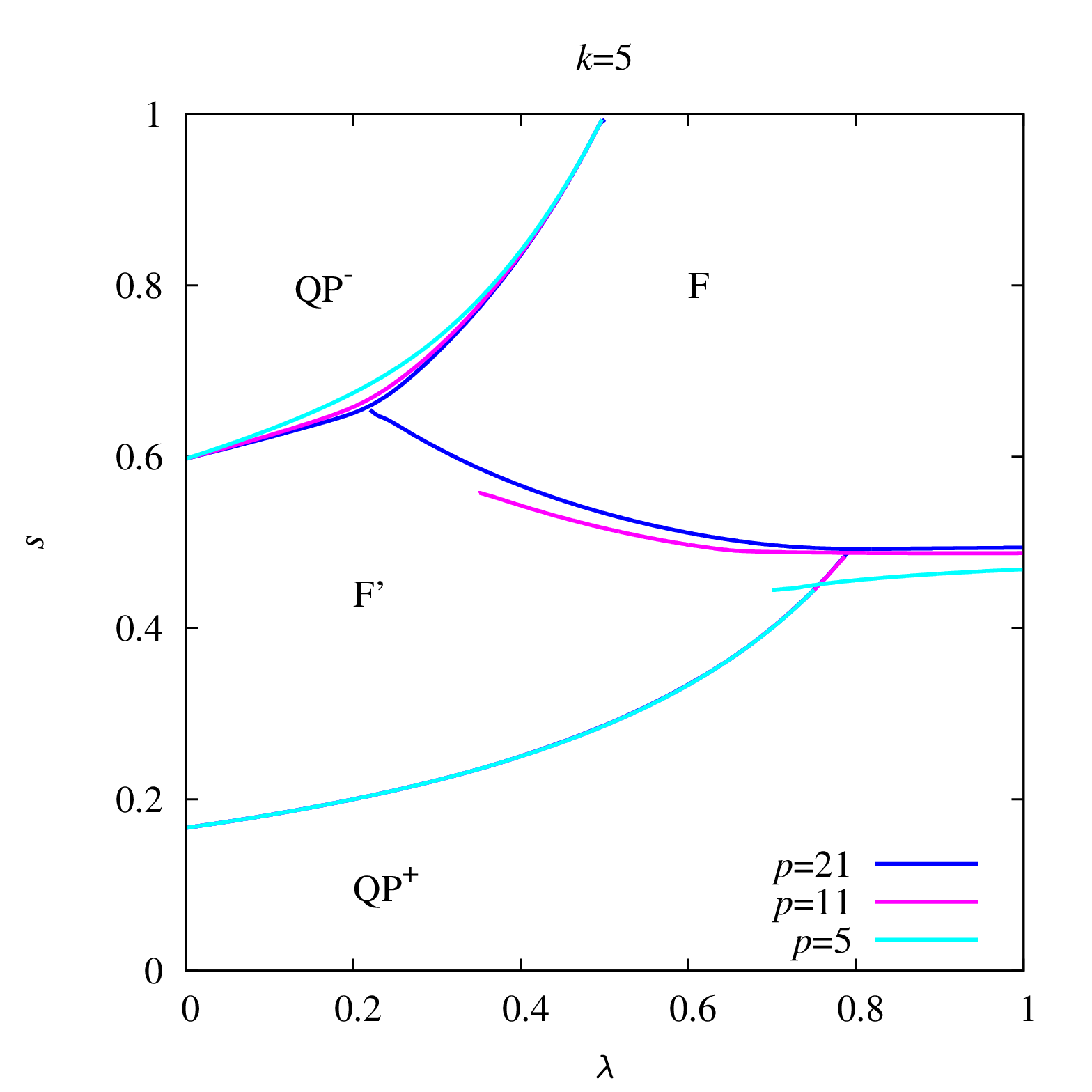}
\caption{Phase diagram  $k=5$. The F'-QP$^+$ transition is of second order, and the other transitions are all of first order.}\label{fig:phase-diag-k5}
 \end{center}
\end{figure}

The phase diagram for finite $p$ is different. Now, there appear regions where
first-order transitions disappear, leaving more space for annealing
trajectories. We display the corresponding diagrams in figures
\ref{fig:phase-diag-k2}, \ref{fig:phase-diag-k3}, \ref{fig:phase-diag-k4} and
\ref{fig:phase-diag-k5} for $k=2$, 3, 4 and 5, respectively. Again, the shape
of the phase diagram strongly depends on whether $k$ is even or odd. In the
former, there are only three phases and in the latter the extra QP$^-$ phase
appears. Besides, the higher $k$ is, the longer is the second-order transition
line.


\begin{figure}
\begin{center}
 \includegraphics[angle=270,width=0.45\columnwidth,trim=0 0 0
   0]{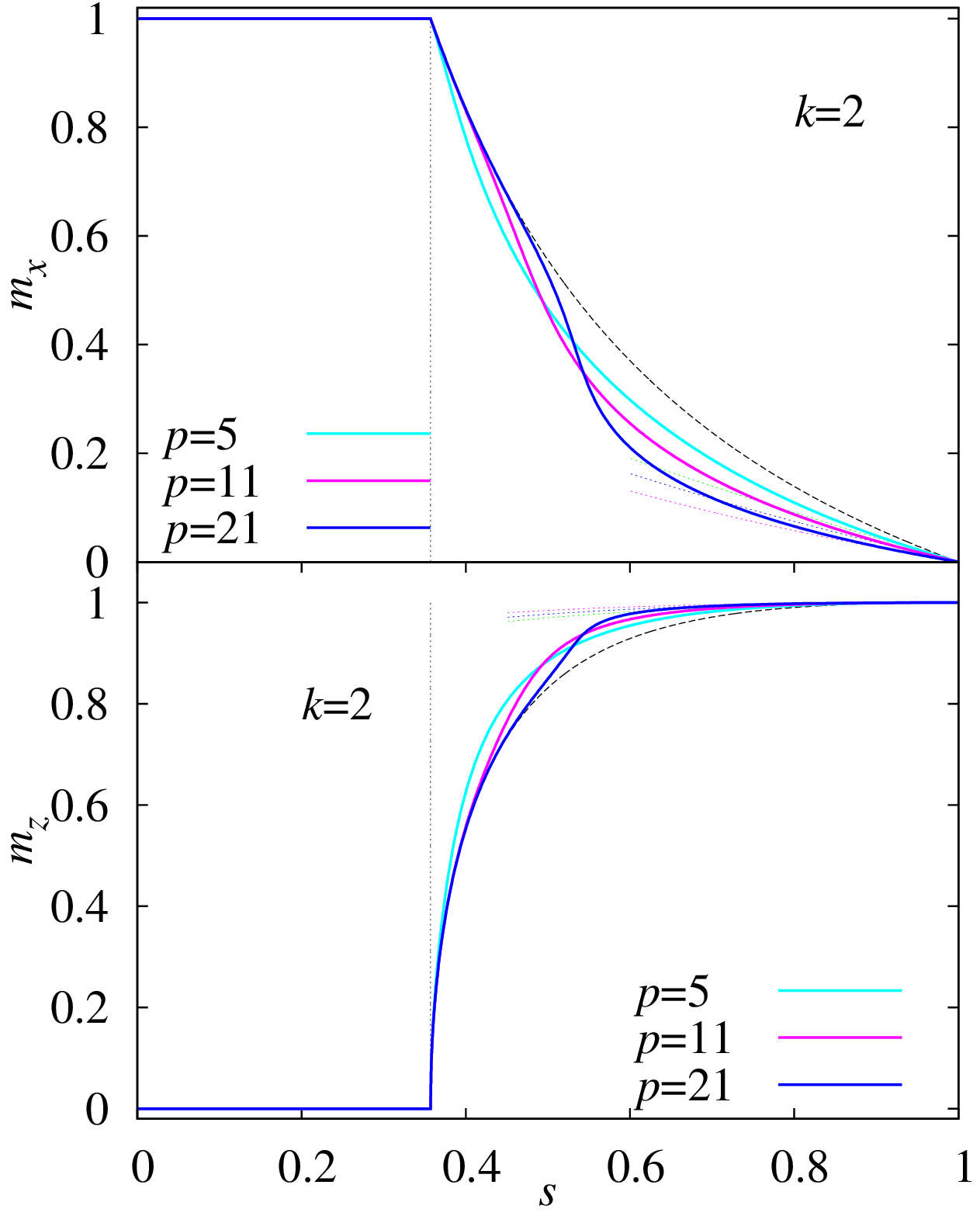}
 \includegraphics[angle=270,width=0.45\columnwidth,trim=0 0 0 0]{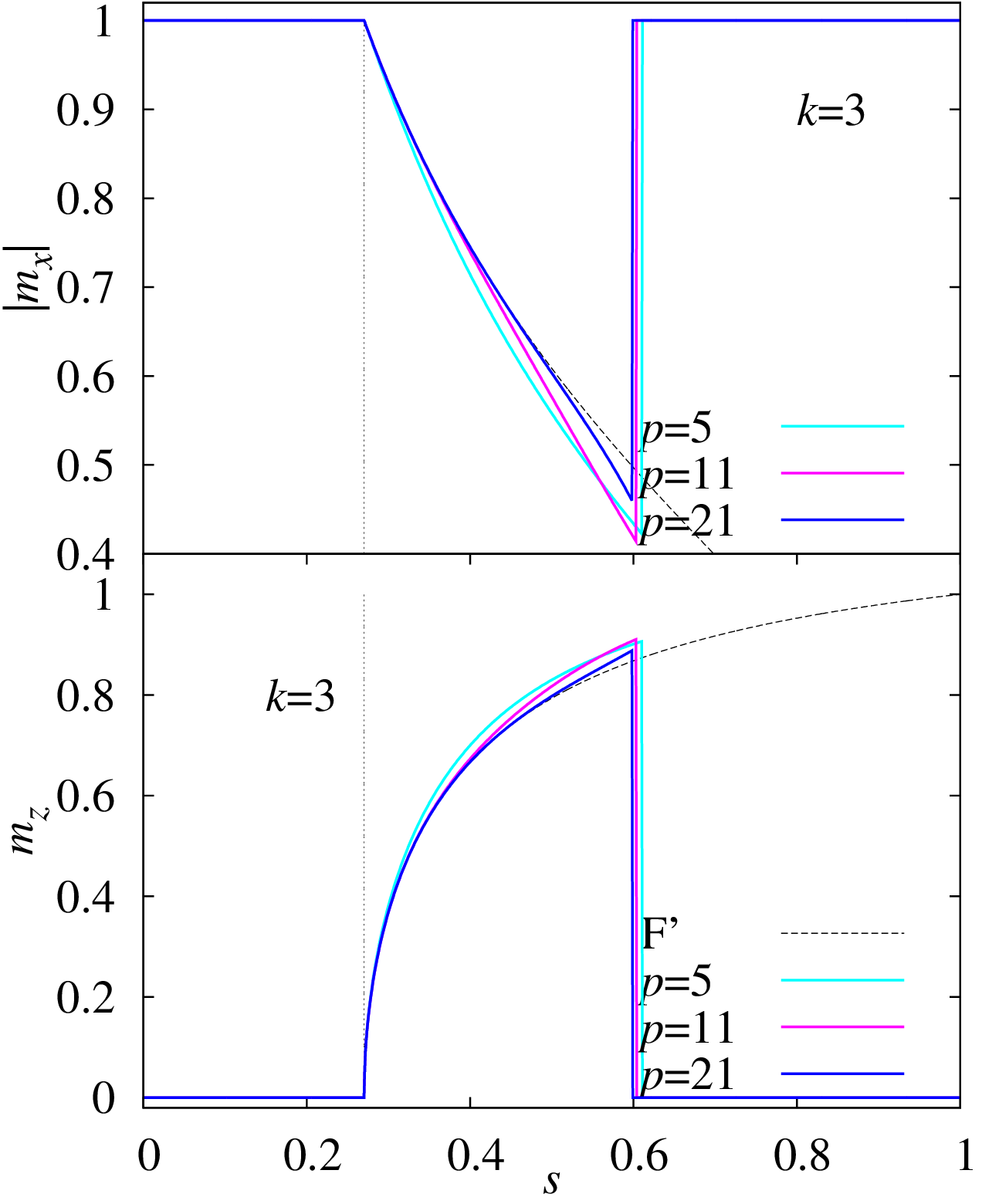}
\caption{Magnetization obtained with the semi-classical approach as a
  function of $s$ for $\lambda=0.1$ and for $k=2$ and 3. The dashed lines
  correspond to the analytical predictions for the QP$^\pm$, F
  \eqref{eq:solFk2} and F' \eqref{eq:solFpCApfin} solutions.}\label{fig:magk2_class_0.1}
 \end{center}
\end{figure}

\begin{figure}
\begin{center}
\includegraphics[angle=270,width=0.45\columnwidth,trim=0 0 0
  0]{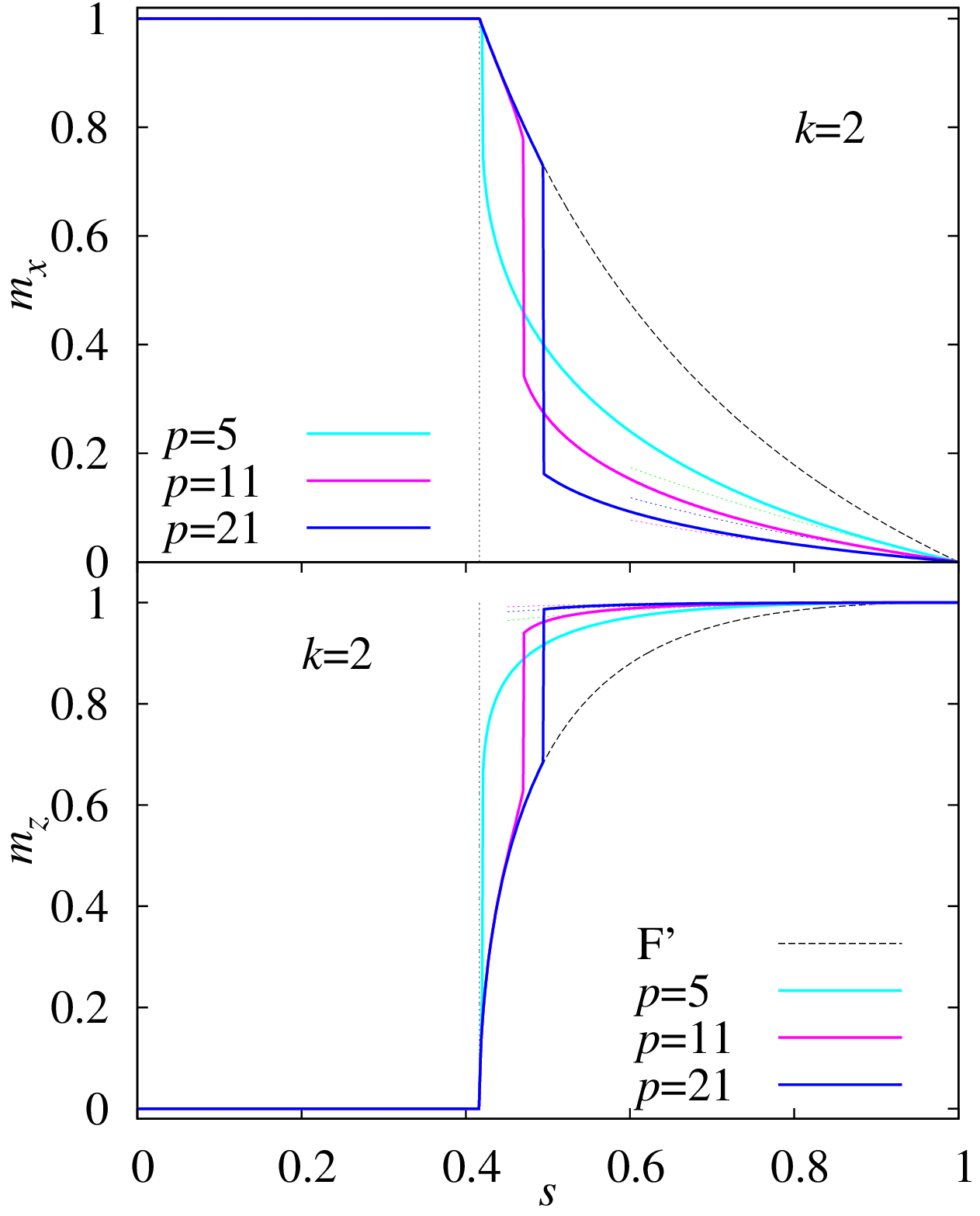}
\includegraphics[angle=270,width=0.45\columnwidth,trim=0 0 0 0]{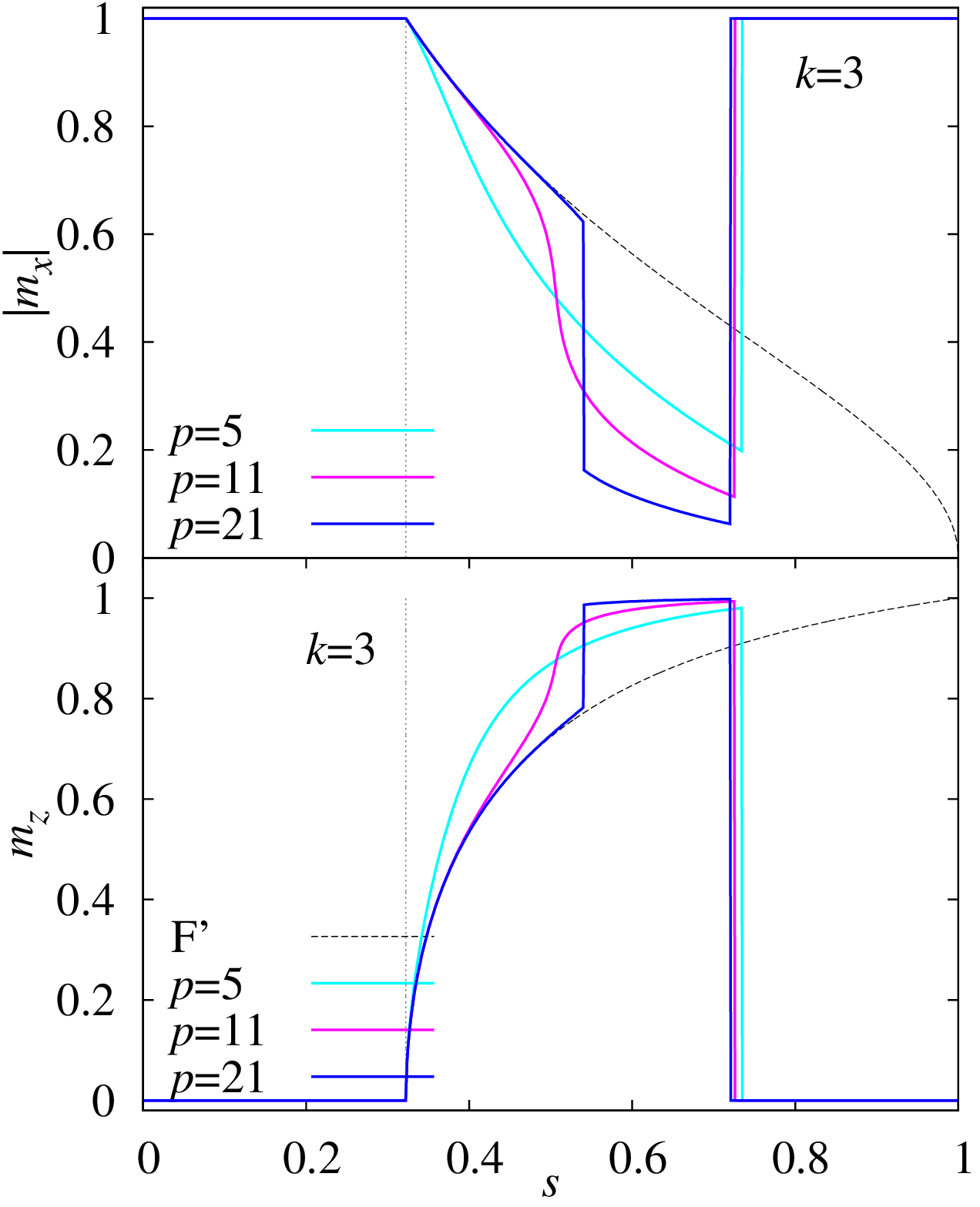}
\caption{Magnetization obtained with the semi-classical  approach as
a  function of $s$ for $\lambda=0.3$ for $k=2$ and 3. The dashed lines
  correspond to the analytical predictions for the QP$^\pm$, F
  \eqref{eq:solFk2} and F' \eqref{eq:solFpCApfin} solutions.}\label{fig:mag_class_0.3}
 \end{center}
\end{figure}

The picture of the ferromagnetic phase for finite $p$ is rather
complicated. When one solves numerically \eqref{eq:eqtheta0} and looks at the
$\theta_0>0$ solutions, the situation is the following: in a wide region, one
finds two possible alternative solutions that look very much alike to the F
and F' phases discussed for the $p\to\infty$ limit. However, near the left and
upper corner in the phase diagram, there is one single ferromagnetic solution
which is neither F nor F' but something intermediate. In fact, for $k$ even,
one can find paths through which the magnetization evolves continuously from
the F' to the F magnetizations without crossing any transition on the way, see
figure \ref{fig:magk2_class_0.1}. However, when $p$ is high and $k$ is odd, transitions between the F and F' phases
cannot be avoided, see figures \ref{fig:phase-diag-k5}, \ref{fig:magk2_class_0.1} and \ref{fig:mag_class_0.3}.

All this effect can be understood quantitatively coming back to
the discussion of the $p\to\infty$ ferromagnetic solutions. Each of the phases
were derived using the assumptions \eqref{eq:limF} for the F phase, and
\eqref{eq:limF'} for the F' phase. Now we discuss the validity of these
approximations for $p$ finite.

We begin with the F phase.  This phase was obtained by introducing
\eqref{eq:limF} in \eqref{eq:eqtheta0}. Since this equality is not strictly
true, we introduce it as an approximation $\sin^{p-2}\theta_0\approx 1$,
thus obtaining a new approximate equation \be
p\,s\,\lambda\cos\theta_0+k\,s\,(1-\lambda)\cos^{k-1}\theta_0-1+s\approx0.
\ee If we assume $\cos\theta_0\ll 1$, the solution is
\be\label{eq:solFk2}
\cos\theta_0\approx\frac{1-s}{s\caja{p\,\lambda+k(1-\lambda)}}, \ee for $k=2$,
and \be \cos\theta_0\approx\frac{1-s}{s\,p\,\lambda}, \ee for $k>2$. That
means, that the F solution found for the $p\to\infty$ limit also appears for
finite $p$ whereas $\cos\theta_0\ll 1$, or \be\label{eq:cociente}
\frac{1-s}{s\,p\,\lambda}\ll 1.  \ee In particular, the smaller this quotient
\eqref{eq:cociente} is, the better approximation the F solution is.  We can
obtain the energies for finite $p$ by introducing this solution in
\eqref{eq:mine}. For $k=2$,
\begin{eqnarray}\nonumber
e^{k=2}_\mathrm{F}(s,\lambda)\approx-s\,\lambda\caja{1-\paren{\frac{1-s}{s\caja{p\,\lambda+2\,(1-\lambda)}}}^2}^{\frac{p}{2}-1}\\+s\,(1-\lambda)\paren{\frac{1-s}{s\caja{p\,\lambda+2\,(1-\lambda)}}}^{k}-(1-s)\paren{\frac{1-s}{s\caja{p\,\lambda+2\,(1-\lambda)}}},
\end{eqnarray}
and for $k>2$
\begin{eqnarray}\label{eq:fFca}\nonumber
e^{k}_\mathrm{F}(s,\lambda)\approx-s\,\lambda\caja{1-\paren{\frac{1-s}{\,s\,p\,\lambda}}^2}^{\frac{p}{2}-1}\\+s(1-\lambda)\paren{\frac{1-s}{s\,p\,\lambda}}^{k}-(1-s)\paren{\frac{1-s}{s\,p\,\lambda}}.
\end{eqnarray}

Next we study the F' solution. We consider the following approximation
\be\label{eq:limitsin} 
p\sin^{p-2}\theta_0\approx0.
\ee As before, if this is a good approximation,
\be\label{eq:solFpCApfin} \cos\theta_0\approx\caja{\frac{1-s}{k\,
    s\,(1-\lambda)}}^{\frac{1}{k-1}} \ee 
is one solution to \eqref{eq:eqtheta0}.  This solution is equal to
the one obtained for $p\to\infty$, 
\eqref{eq:solFpCA}. In other words, at this order of approximation, the
solution is exact at this limit.

We briefly discuss the range of validity of this F' solution
\eqref{eq:solFpCApfin} for $p$ finite. The approximation \eqref{eq:limitsin} is
valid for small values of $\theta_0$. With this idea we expand separately
the two terms in \eqref{eq:eqtheta0} around $\theta_0=0$, the
first term being
\begin{eqnarray}\nonumber
\sin^{p-2}\theta_0\cos\theta_0&=&p\,s\,\lambda\,\theta_0^{p-2}\caja{1-\frac{p+1}{6}\theta_0^2+O(\theta_0^{4})},
\end{eqnarray}
and the second term
\begin{eqnarray}\nonumber
k\,s\,(1-\lambda)\cos^{k-1}\theta_0-1+s
\\=k\,s\,(1-\lambda)\caja{1-\frac{k-1}{2}\theta_0^2+O(\theta_0^4)}-1+s.
\end{eqnarray}
The dependency on $\theta_0$ in the first term becomes irrelevant when
$p>3$, thus recovering the F' solution \eqref{eq:solFpCApfin}. When $p=3$, the
lowest power of $\theta_0$ appears in the first term,
leading to a different ferromagnetic solution, but not the F'. Clearly, the
higher $p$ (and the smaller $\theta_0$) is, the better is approximation
\eqref{eq:limitsin}.

In general, for intermediate values of $s$ and $\lambda$, the higher $p$ is,
the more exact the two ferromagnetic solutions, F and F', are. Then, since
both approximations represent opposite cases in the value of $m^x$ (or $m^z$),
a new first-order transition between both phases will appear on the line when
their two free energies become equal. However, for low values of $p$, or
alternatively for $s\to1$ or $0$, there will only be one ferromagnetic
solution, somewhere in between these two F and F' phases. This idea is well
ilustrated in figures \ref{fig:magk2_class_0.1} and \ref{fig:mag_class_0.3},
where both the numerical solution to \eqref{eq:eqtheta0} and the analytical
predictions  \eqref{eq:solFk2} and \eqref{eq:solFpCA} are displayed.

This has straightforward consequences on the performance of the quantum
annealing algorithm: the higher $p$ is, the narrower will be the region where
annealing paths can avoid a first-order transition. In the limit of
$p\to\infty$, as was discussed before, there will be only one possible path,
but not effective as quantum annealing.

Concerning the transitions between the QP and ferromagnetic phases, we can
distinguish two kinds of transitions. First of all, the transitions between
the F and QP$^\pm$ phases will be first order, since the F phase is
characterized by a high value of $m^z$ whereas the paramagnetic solution has
$m^z=0$.  On the other hand, there is another transition between the F' and QP
phases that lies on the line where their two free energies become equal,
i.e. $s=1/[1+k(1-\lambda)]$. On this line, $m^x=1$ ($m^z=0$) for the two
phases. Furthermore, the F' solution is exact for $m^x=1$. Since the
magnetizations are continuous on this line, the transition between F' and QP is
of second order. Besides, it can be checked that there is a wide range of
this line where $e_\mathrm{F'}<e_\mathrm{F}$. Thus, this phase is the stable
one in the ferromagnetic phase. This second-order transition does not hamper
the QA performance and gives us a way to avoid the F-QP phase transition that
appeared when using the traditional QA approach.  It is important to point out
that this second-order transition appears for any value of $k$.

In \ref{sec:SA}, we describe a different, quantum-mechanical method to derive the same results.

\section{Energy gap}
As discussed in Introduction, the efficiency of the QA algorithm is closely
related to the behavior of the gap between the ground and first excited
states. As usual, this gap can be computed by direct diagonalization of the
problem Hamiltonian \eqref{eq:H}.  Indeed, since the total spin $\V{S}$ is
 conserved during the evolution, the dimension of the problem is $N+1$. That
 means that the diagonalization matrixes grow polynomially with the system
 size instead of exponentially as for generic quantum problems. However, still 
 computer resources limit this
computation to moderate sizes although such computations are useful
for some purposes~\cite{seki:12,jorg:10a}.   Here, we adopt an
  alternatively approach, this gap can be
computed in the thermodynamic limit $N\to\infty$ by the method described
in~\cite{filippone:11}. The main idea is to extend the semi-classical scheme
for the ground state by the consideration of quantum fluctuations around the
classical ground state. It is important to point out that this method can
only be applied in the case of finite gaps in the thermodynamic limit, as it
is the case away from the transition points themselves. In case of exponentially small ones, other methods such as
instantonic or WKB methods should be used~\cite{jorg:10a,bapst:12}.

It is most convenient to rotate the system by an angle $\theta_0$ around the
$y$ axis in order to bring the $x$ axis parallel to the semi-classical
magnetization, i.e.
\begin{equation}
\paren{\begin{array}{c} S_x\\S_y\\S_z \end{array}}=\paren{\begin{array}{ccc}
    -\sin{\theta_0}&0&\cos{\theta_0}\\0&1&0\\\cos{\theta_0}&0&\sin{\theta_0}\end{array}}\paren{\begin{array}{c}
    \tilde S_x\\\tilde S_y\\\tilde S_z \end{array}}.
\end{equation}
We rewrite the Hamiltonian \eqref{eq:Hnew} in terms of these new variables
$\tilde S^\alpha$, obtaining
\begin{eqnarray}\label{eq:Hrot}\nonumber
&\hat{H}(s,\lambda)=-s\,\lambda\, N\caja{\frac{2}{N}\paren{\cos\theta_0\,\tilde
    S_x+\sin\theta_0\,\tilde S_z}}^p&\\\nonumber&+s\,(1-\lambda)\,N\,\caja{\frac{2}{N}\paren{-\sin\theta_0\,\tilde S_x+\cos\theta_0\,\tilde S_z}}^k&\\&-2\,(1-s)\paren{-\sin\theta_0\,\tilde S_x+\cos\theta_0\,\tilde S_z}.&
\end{eqnarray}

Now, we add quantum fluctuations to the system by means of the
Holstein-Primakoff transformation
\begin{eqnarray}\label{eq:HP}
\tilde S_z=\frac{N}{2}-a^\dagger a,\,\,\,\,\,\,&\tilde S_+=\displaystyle(N-a^\dagger
a)^{1/2} a=\tilde S_-^\dagger,
\end{eqnarray}
where $a$ is a boson annihilation operator that satisfies $[a,a^\dagger]=1$.
When quantum fluctuations are small relative to the classical state, i.e. for $N\gg\mean{a^\dagger a}$, we can use a simpler expression \be\tilde
S_x\approx\displaystyle\frac{\sqrt{N}}{2}(a+a^\dagger).\ee We introduce these
transformations into the Hamiltonian \eqref{eq:Hrot} and expand the three
different terms in powers of $1/N$. Thanks to the previous rotation, the
coefficient in $1/\sqrt{N}$ vanishes. We keep terms up
to $1/N$ and group together all the coefficients with the same power of $N$.
The result is 
\be\label{eq:Hhp}
H(\gamma,\delta)=N\,e+\gamma+\gamma\caja{(a^\dagger)^2+a^2}+\delta a^\dagger
a.  \ee 
The term for $N^1$ is nothing but the ground energy obtained before in
\eqref{eq:mine}, \be
e\equiv-s\lambda\sin^p\theta_0+s(1-\lambda)\cos^k\theta_0-(1-s)\cos\theta_0.
\ee 
The coefficients $\delta$ and $\gamma$ are given as
\begin{eqnarray}\nonumber
\delta&\equiv&-s\,\lambda\caja{p(p-1)\sin^{p-2}\theta_0\cos^2\theta_0-2\,p\,\sin^p\theta_0}\\&+&s\,(1-\lambda)\caja{k(k-1)\sin^{2}\theta_0\cos^{k-2}\theta_0-2\,k\,\cos^k\theta_0}+2(1-s)\cos\theta_0,
\end{eqnarray}
and 
\begin{eqnarray}
\gamma\equiv-\frac{s\,\lambda\,p(p-1)}{2}\sin^{p-2}\theta_0\cos^2\theta_0+s\,(1-\lambda)
  \frac{k(k-1)}{2}\sin^{2}\theta_0\cos^{k-2}\theta_0.
\end{eqnarray}
We need to diagonalize this Hamiltonian in order to compute the first excited
state by the Bogoliubov transformation
\begin{eqnarray}
a=\cosh\frac{\Theta}{2}\,b+\sinh\frac{\Theta}{2}\,b^\dagger,&\displaystyle
a^\dagger=\cosh\frac{\Theta}{2}\,b^\dagger+\sinh\frac{\Theta}{2}\,b,
\end{eqnarray}
where $b$ is a new bosonic annihilation operator satisfying $[b,b^\dagger]=1$.
Using this transformation, we can eliminate the coefficient of
$\caja{(b^\dagger)^2+b^2}$ by choosing the angle $\Theta$ as
\begin{eqnarray}\nonumber
 \tanh\Theta=-\frac{2\gamma}{\delta}\equiv\epsilon.
\end{eqnarray}
With this choice, the Hamiltonian can be written as
\be\label{eq:HB}
H(\gamma,\delta)=N\,e+\gamma+\frac{\delta}{2}\paren{\sqrt{1-\epsilon^2}-1}+\Delta\ b^\dagger b,
\ee
with \be\label{eq:Delta}
\Delta=\delta\sqrt{1-\epsilon^2}.
\ee
The Hamiltonian is diagonal in
$b^\dagger b$. The energy gap in the $N\to\infty$ limit between the ground 
and first excited states is  $\Delta$.

\begin{figure}
\begin{center}
 \includegraphics[angle=270,width=0.45\columnwidth,trim=10 20 10
   20]{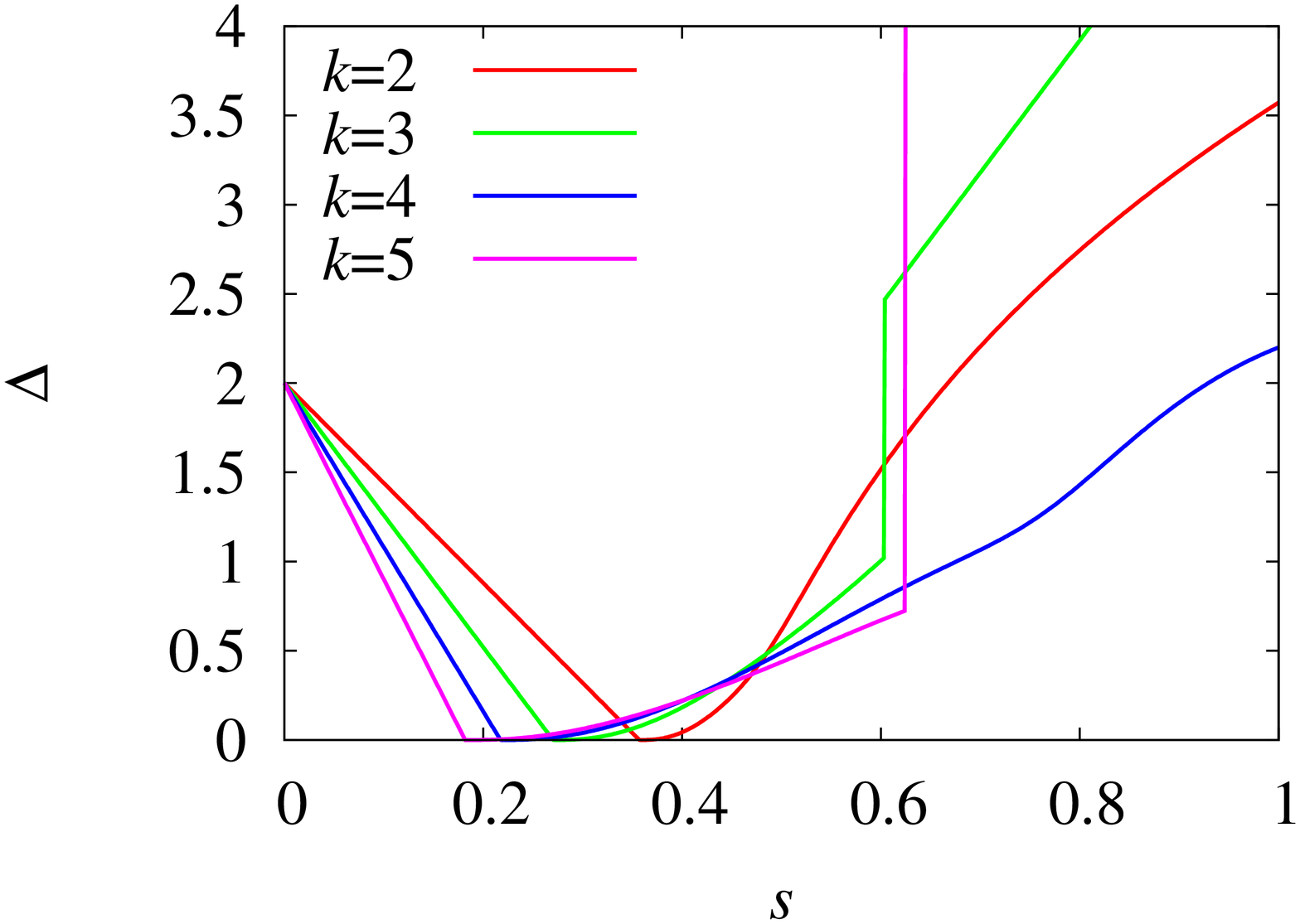}
 \includegraphics[angle=270,width=0.45\columnwidth,trim=10 20 10
   20]{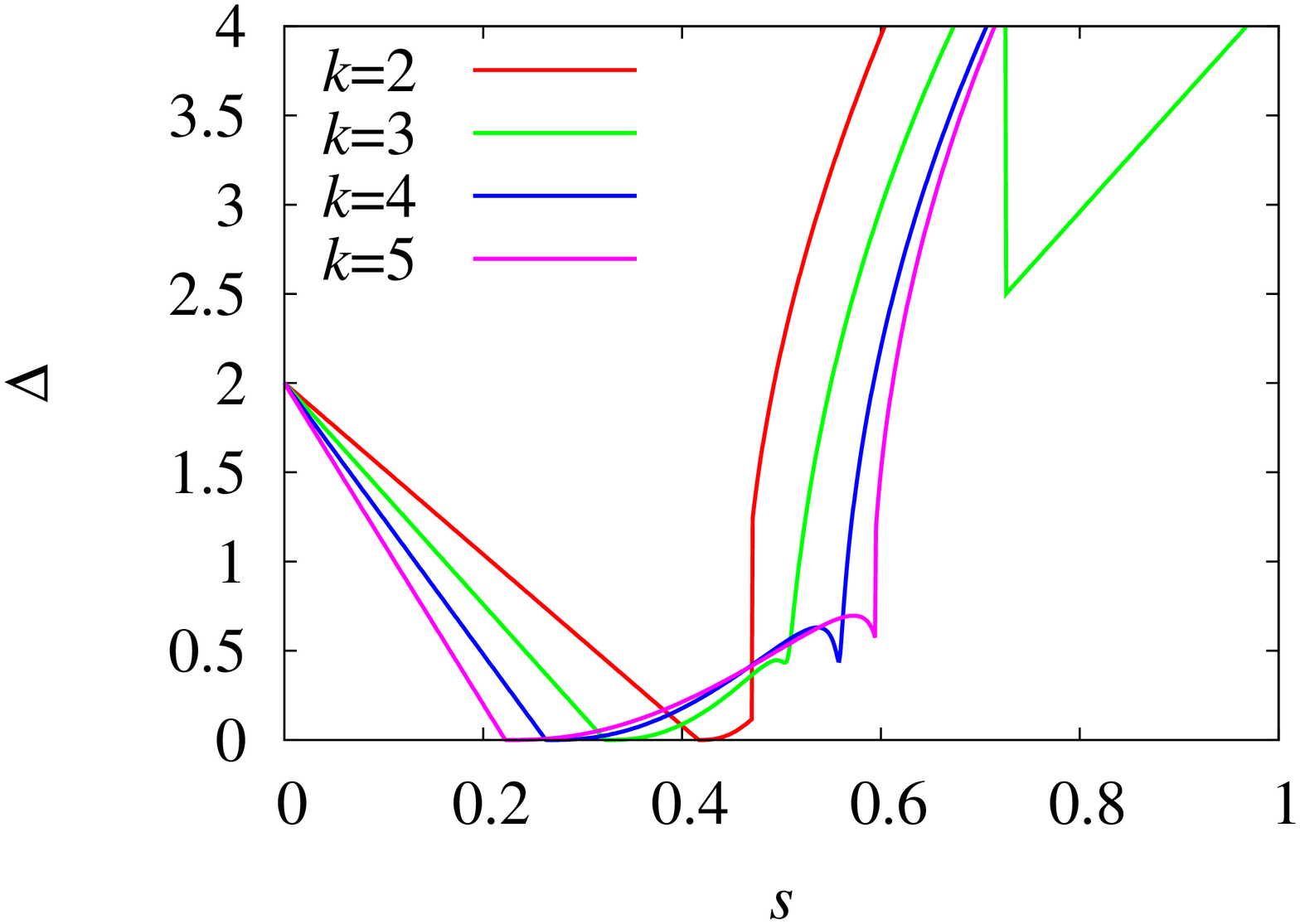}
\caption{Energy gap for $p=11$ as a function of $s$ for $\lambda=0.1$ (left) and
  $\lambda=0.3$ (right) for several values of $k$.}\label{fig:overlap}
 \end{center}
\end{figure}

Using the values $\theta_0$ previously obtained solving
 \eqref{eq:eqtheta0}, we can compute the energy gap for our
system. We show the data for $p=11$ and $\lambda=0.1$ and $0.3$  for
different values of $k$ in figure \ref{fig:overlap}. As was suggested in the
magnetization data in the previous section for $\lambda=0.1$ (figures
\ref{fig:phase-diag-k2} to
\ref{fig:phase-diag-k5}), no first-order
transition F-F' is observed through the energy gap. The gap vanishes continuously on the
second-order transition line but present no further jumps later, but the ones
related to the F-QP$^-$ that always take place in the odd-$k$ cases. On the
contrary, when $\lambda=0.3$, the jumps in the gap appear for all the $k$'s at
the place where we observed the F-F' transition before.

In the thermodynamic limit, the gap vanishes at a single point of first-order transition and remains finite away from this point.  The single point of vanishing gap is hard to see by the present method, which results in an apparent simple jump in the gap at a first-order transition as seen in figure \ref{fig:overlap}.

\section{Overlap of the ground-state wave functions}
It has been suggested in~\cite{bapst:12} that
the reason for the antiferromagnetic interaction, the $k=2$ case in
 \eqref{eq:Vk}
introduced by Seki and Nishimori in \cite{seki:12}, to work better than the
transverse field interaction only is related to the large overlap
between the ground states of the  Hamiltonians $\hat{V}_{k=2}$ and $\hat{H}_0$.
In this section, we will discuss the properties of these
different states, concluding that, even thought the overlap is important,
it is not the decisive factor that makes the strategy to succeed.

The ground state of $\hat V_\mathrm{TF}$ is the one where
all the spins are aligned along the $x$ axis,
$\ket{\phi_\mathrm{TF}}=\otimes_{i=1}^N\ket{\uparrow}_i^x$. If we denote the
ground state of $\hat{H}_0$, as
$\ket{\phi_0}=\otimes_{i=1}^N\ket{\uparrow}_i^z$,  the overlap between $\ket{\phi_\mathrm{TF}}$
and $\ket{\phi_0}$ decreases exponentially with $N$ as $2^{-N}$, as can easily
be seen from the elementary relation $\ket{\uparrow}_i^x=\paren{\ket{\uparrow}_i^z+\ket{\downarrow}_i^z}/\sqrt{2}$.

The overlap computation becomes a little more complicated for the ground state
of $\hat{V}_k$. The ground state for this term depends on the value of
$k$. Indeed, if $k$ is odd, the ground state is the one where
all the spins are aligned along the $x$ axis, but towards the negative
direction, i.e.  $\otimes_{i=1}^N\ket{\downarrow}_i^x$. Then, the overlap with
$\ket{\phi_0}$ for the $k$ odd case will be exponentially suppressed as
$2^{-N}$ as in the case of $\hat{V}_\mathrm{TF}$. Thus, the argument
in~\cite{bapst:12} does not apply directly since we can avoid first-order
transitions even in this case of $k$ odd, in spite of the very small overlap
of the ground state for $\hat{H}_0$ and $\hat{V}_k$.

The ground state for the $k$ even case needs some care to be analyzed. We compute it in
 \ref{sec:Vkground}. We show there that  that the overlap is  indeed higher
for  $k$ even. The antiferromagnetic interations is a particular case, $k=2$. In
fact, the overlap displays an algebraic decay as the system size increases, i.e. $\sim 1/\sqrt{N}$.

We conclude that the overlap is not the main ingredient that makes
the present method to succeed. 
\section{Conclusions}
We have analyzed the reason for the failure of the traditional annealing with
a transverse-field term in the infinite-range ferromagnetic $p$-spin model.
We have shown that it is possible to find annealing trajectories that avoid
the crossing of first-order transitions thanks to the introduction of a second
driver term in the problem, which may be due to the multiple spin flips in the $z$-basis caused by the second term as was the case in~\cite{suzuki:07}.  This additional term favors the appearance of a
second-order transition that does not hamper the annealing performance.  A
whole family of possible candidates has been studied and we conclude that the
solution to the problem presented by Seki and Nishimori~\cite{seki:12} is a
special case of a more general additional quantum term. The main properties of these additional terms have also been
discussed with the conclusion that the properties of the ground states of the
diverse terms in the Hamiltonian are not a decisive factor to make the quantum
annealing fail or succeed.

\section{Acknowledgments}
Beatriz Seoane was supported by the FPU
program (MECD, Spain).

\vspace{0.7cm}
\begin{footnotesize}
{\em Note added in Proof:} After the submission of the manuscript, it was
pointed out that our Hamiltonians \eqref{eq:H0} and \eqref{eq:Vk} were a special case of what was
studied in \cite{denouden:76,denouden:76b,perk:77}.
\end{footnotesize}

\appendix
\section{Analysis with the Suzuki-Trotter formula}\label{sec:SA}
We investigate the properties of $\hat H(s,\lambda)$, the phase diagram in
particular, using the decomposition formula \cite{suzuki:76} and the static
approximation. This approach, although quantum, leads to the same results as
the semi-classical method described in section \ref{sec:CA}. The method here
is analogous to the one explained in detail in \cite{seki:12,jorg:10a}, but we
leave the power $k$ as a free parameter in all the calculus. The purpose of
this appendix is to confirm consistency between the method of the main text
and that in \cite{seki:12,jorg:10a}.

The starting point is the partition function, \be\label{eq:Z} Z=\Tr
e^{-\beta\hat{H}(s,\lambda)}.  \ee We use the  decomposition
formula to express it as
\begin{eqnarray}\label{eq:Z0}\nonumber
Z&=&\lim_{M\to\infty}Z_M\equiv\lim_{M\to\infty}\Tr
\lazo{e^{-\frac{\beta}{M}
    s\lambda\hat{H}_0}e^{-\frac{\beta}{M}\caja{s\,(1-\lambda)\hat{V}_{\mathrm{AFF}}+(1-s)\hat{V}_{\mathrm{TF}}}}}^M\\\nonumber
&=&\lim_{M\to\infty}\sum_{\{\sigma^z\}}\bra{\{\sigma^z\}}\left\{\exp\caja{\frac{\beta
      s\lambda N}{M}\paren{\frac{1}{N}\sump\hsiz}^p}\right.\\\nonumber
&&\left.\times\exp\caja{-\frac{\beta s\,(1-\lambda) N}{M}\paren{\frac{1}{N}\sump\hsix}^k+\frac{\beta (1-s) }{M}\sump\hsix}\right\}^M\ket{\{\sigma^z\}},\\
\end{eqnarray}
where $\sum_{\{\sigma^z\}}$ refers to the summation over all the $2^N$
possible spin configurations in the $z$ basis, and
$\ket{\{\sigma^z\}}\equiv\otimes_{i=1}^N\ket{\siz}$.  

We introduce $M$ closure relations, each one labeled by $\alpha(=1,\ldots,M)$,
\be \hat{\mathbb{I}}(\alpha)\equiv
\sum_{\{\sigma^z(\alpha)\}}\ket{\{\sigma^z(\alpha)\}}\bra{\{\sigma^z(\alpha)\}}\times\sum_{\{\sigma^x(\alpha)\}}\ket{\{\sigma^x(\alpha)\}}\bra{\{\sigma^x(\alpha)\}},
\ee just before the $\alpha$th exponential operator involving $\hsix$ in
\eqref{eq:Z0}. The trace over the product of quantum operators is thus reduced
to the product of numbers that commute and can be reordered,
\begin{eqnarray}\label{eq:ZM}\nonumber
Z_M&=&\prod_{\alpha=1}^M
  \sum_{\{\sigma^z(\alpha)\}}\sum_{\{\sigma^x(\alpha)\}} 
\exp\caja{\frac{\beta s \lambda N }{M}\paren{\frac{1}{N}\sump\siz(\alpha)}^p}\\\nonumber
&&\times\exp\caja{-\frac{\beta s (1-\lambda) N
  }{M}\paren{\frac{1}{N}\sump\six(\alpha)}^k+\frac{\beta (1-s)
  }{M}\sump\six(\alpha)}\\
&&\times\prod_{i=1}^N\left\langle\siz(\alpha)\right.\ket{\six(\alpha)}\left\langle\six(\alpha)\right.\ket{\siz(\alpha+1)},
\end{eqnarray}
where $\ket{\siz(M+1)}\equiv\ket{\siz(1)}$.

We write the product in terms of the total $x$ and $z$ magnetizations in each
copy of the system, i.e.  $m^x(\alpha)\equiv\frac{1}{N}\sump\six(\alpha)$ and
$m^z(\alpha)\equiv\frac{1}{N}\sump\siz(\alpha)$, using the integral definition
of the delta distribution \be
f\paren{\frac{1}{N}\sump\sigma_i(\alpha)}=\int\mathrm{d}m\,\delta\paren{m(\alpha)-\frac{1}{N}\sump\sigma_i(\alpha)}f\paren{m(\alpha)}.
\ee After a few simplifications, we introduce the static approximation to
remove the $\alpha$ dependence of the magnetizations. Under this
approximation, we can compute the $M\to\infty$ limit using again the
decomposition formula.  The partition function \eqref{eq:Z} then reduces to
\be
Z=\int\mathrm{d}m^z\,\mathrm{d}m^x\,\exp\caja{-N\beta\,f(\beta,s,\lambda;m^z,m^x)},
\ee where $f(\beta,s,\lambda;m^z,m^x)$ is the pseudo free-energy defined as
follows:
\begin{eqnarray}\label{eq:freebeta}\nonumber
f(\beta,s,\lambda;m^z,m^x)=(p-1)\, s\,
    \lambda(m^z)^p-\, (k-1)\, s\,
      (1-\lambda)(m^x)^k\\-\frac{1}{\beta}\log\lazo{2 \cosh \beta\sqrt{\caja{p\, s\,\lambda\, (m^z)^{p-1}}^2+\caja{1-s-s\,(1-\lambda)\,k\,(m^x)^{k-1}}^2}}.
\end{eqnarray}
Again, one can apply the saddle-point method, obtaining two self-consistent equations
for the two magnetizations,
\begin{eqnarray}\label{eq:mz}
m^z&=&\frac{p\, s\,\lambda\, (m^z)^{p-1}}{\sqrt{\caja{p\, s\,\lambda\,
      (m^z)^{p-1}}^2+\caja{1-s-s\,(1-\lambda)\,k\,(m^x)^{k-1}}^2}}\\\nonumber
&\times& \tanh \beta\sqrt{\caja{p\, s\,\lambda\,
      (m^z)^{p-1}}^2+\caja{1-s-s\,(1-\lambda)\,k\,(m^x)^{k-1}}^2},\\\label{eq:mx}
m^x&=&\frac{1-s-s\,(1-\lambda)\,k\,(m^x)^{k-1}}{\sqrt{\caja{p\, s\,\lambda\,
      (m^z)^{p-1}}^2+\caja{1-s-s\,(1-\lambda)\,k\,(m^x)^{k-1}}^2}}\\\nonumber
&\times& \tanh \beta\sqrt{\caja{p\, s\,\lambda\,
      (m^z)^{p-1}}^2+\caja{1-s-s\,(1-\lambda)\,k\,(m^x)^{k-1}}^2}.
\end{eqnarray}

In this work we are only interested in the purely quantum transitions, not
in the thermodynamical ones. For this reason, and with the sake of
simplification, we remove the dependence of physical quantities on $\beta$ from now on by
considering the low-temperature limit,  $\beta\to\infty$.
In this limit, if $\caja{p\, s\,\lambda\,
  (m^z)^{p-1}}^2+\caja{1-s+s\,(1-\lambda)\,k\,(m^x)^{k-1}}^2\neq 0$, the
hyperbolic tangent in  \eqref{eq:mz} and \eqref{eq:mx} tends to unity,
and thus the self consistent equations simplify
\begin{eqnarray}\label{eq:mzbetainfty}
m^z&=&\frac{p\, s\,\lambda\, (m^z)^{p-1}}{\sqrt{\caja{p\, s\,\lambda\,
      (m^z)^{p-1}}^2+\caja{1-s-s\,(1-\lambda)\,k\,(m^x)^{k-1}}^2}},\\\label{eq:mxbetainfty}
m^x&=&\frac{1-s-s\,(1-\lambda)\,k\,(m^x)^{k-1}}{\sqrt{\caja{p\, s\,\lambda\,
      (m^z)^{p-1}}^2+\caja{1-s-s\,(1-\lambda)\,k\,(m^x)^{k-1}}^2}}.
\end{eqnarray}
The magnetization lies on the unit radius circumference, i.e.
$(m^x)^2+(m^z)^2=1$. This result agrees with the approach in section
\ref{sec:CA}, where the magnetization was a unit vector constrained to the
$XZ$ plane.  The pseudo free energy \eqref{eq:freebeta} becomes
\begin{eqnarray}\label{eq:free}\nonumber
&f(\beta,s,\lambda;m^z,m^x)=(p-1)\, s\,
    \lambda(m^z)^p-(k-1)\, s\,
      (1-\lambda)(m^x)^k&\\&-\sqrt{\caja{p\, s\,\lambda\, (m^z)^{p-1}}^2+\caja{1-s-s\,(1-\lambda)\,k\,(m^x)^{k-1}}^2}.&
\end{eqnarray}
Equations \eqref{eq:mzbetainfty} and \eqref{eq:mxbetainfty} have 
ferromagnetic (F) solutions with $m^z>0$ and quantum paramagnetic (QP) ones
satisfying $m^z=0$ and $m^x\neq0$. Let us begin with the latter case.
\subsection{Paramagnetic solutions}
Substituting $m^z=0$ in \eqref{eq:mxbetainfty}, we get
\begin{equation}\label{eq:mxQP}
m^x=\frac{1-s-k\,s\,(1-\lambda)(m^x)^{k-1}}{|1-s-k\,s\,(1-\lambda)(m^x)^{k-1}|},
\end{equation} 
which leads to $m^x=\pm 1$. The solution $m^x=-1$ is obtained if the
numerator in \eqref{eq:mxQP} is negative, that is, if $1-s-k\,s\,(1-\lambda)(-1)^{k-1}< 0$, which, in the range of parameters
$0\le s\le 1$ and $0\le \lambda\le 1$ considered, can only be satisfied if $k$ is 
  odd and in the region $1/[1+k(1-\lambda)]< s\le 1$. This phase is
precisely the $\mathrm{QM}^-$ phase discussed in the text. Its free energy is 
\begin{equation}\label{eq:freeQPneg}
  f_{\mathrm{QP}^{-}}(s,\lambda)=1-2s+s\lambda, \end{equation} which coincides
with equation \eqref{eq:fQPmca}.

The other quantum paramagnetic solution with $m^x=+1$ (the $\mathrm{QP}^+$
phase) can be satisfied only if the numerator is positive, i.e. if
$1-s-k\,s\,(1-\lambda)\geq 0$, which can be fulfilled for any value of $k$ as
long as $s$ lies in the region $0\leq s\leq1/[1+k(1-\lambda)]$. The free
energy of this phase is \begin{equation}\label{eq:feQP}
  f_{\mathrm{QP}^{+}}(s,\lambda)=-1+2s-s\lambda, \end{equation}
and is also equal to \eqref{eq:fQPpca}.

There is still one additional paramagnetic solution. In order to obtain it, we
need to come back to the discussion about the $\beta\to\infty$ limit.  The
hyperbolic tangent in  \eqref{eq:mz} and \eqref{eq:mx} could tend to a
finite value in the $\beta\to\infty$ limit, as long as the term in the square
root vanishes. Mathematically,\footnote{\label{fn:mxneg} In the $k$-odd case, the limit 
$$m^z\to0,\,\,m^x\to-\displaystyle\caja{\frac{1-s}{k\,s\,(1-\lambda)}}^\frac{1}{k-1}$$
also makes the square root in \eqref{eq:limtanh} vanish, but it leads to a positive
free energy in \eqref{eq:freeQP2}, and thus it is not relevant.}
\begin{equation}\label{eq:limtanh} \lim_{\beta\to\infty} \tanh \beta\sqrt{\caja{p\,
    s\,\lambda\, (m^z)^{p-1}}^2+\caja{1-s-s\,(1-\lambda)\,k\,(m^x)^{k-1}}^2}=
\tanh c,  \end{equation} when
\begin{eqnarray}\label{eq:limQP2}
m^z\to0,&m^x\to\displaystyle\caja{\frac{1-s}{k\,s\,(1-\lambda)}}^\frac{1}{k-1}.
\end{eqnarray}

In order to find a non-trivial solution, it is also necessary in this limit
that $m^z$ tends to zero faster than the bracketed term of $m^x$ in
\eqref{eq:mx}, i.e.  
\begin{equation}\label{eq:condQP2}
\frac{p\,s\,\lambda (m^z)^{p-1}}{1-s-k\,s\,(1-\lambda)(m^x)^{k-1}}\to 0. \end{equation}
Under these assumptions,  \eqref{eq:mz} and \eqref{eq:mx} imply $m^z=0$ and
$m^x=\tanh c$, where $\tanh c=[(1-s)/k\,s\,(1-\lambda)]^\frac{1}{k-1}$, in order to
be consistent with the limit \eqref{eq:limQP2}. This correspondence
determines the region in the space where this phase can appear. In fact, as any hyperbolic
tangent, $|\tanh c|\le 1$, which is  true only if $1/[1+k\,(1-\lambda)]\le
s\le 1$. Besides, the condition \eqref{eq:condQP2} forces $p>3$.\footnote{
Indeed, using $(m^x)^2+(m^z)^2=\tanh^2
c=[(1-s)/k\,s\,(1-\lambda)]^\frac{2}{k-1}$ and computing the limit
\eqref{eq:limQP2} when $m^x\to\tanh c$, one can check that it vanishes only  as
long as $p>3$.
}

Since the magnetization in the $z$ direction vanishes, we call this phase
QP2. The free energy is obtained with  \eqref{eq:freebeta},
\begin{equation}\label{eq:freeQP2}
f_{\mathrm{QP2}}(s,\lambda)=-\frac{k-1}{k}\caja{\frac{1-s}{k\,s\,(1-\lambda)}}^\frac{1}{k-1}(1-s).
\end{equation}

This last phase was not predicted by the semi classical approach. However, we will
see below that it is irrelevant to the problem, since the F' phase has
always a smaller value of the free energy.

\subsection{Ferromagnetic solutions}
We next consider the possible solutions with $m^z>0$.\footnote{No negative
  value for $m^z$ can satisfy \eqref{eq:mzbetainfty} for odd values of $p$.}
As before, the ferromagnetic solutions cannot be computed explicitly for a
given value of $p$ but for certain limiting cases.

The solution $m^z=1$ (and $m^x=0$) is  exact only
on the line $s=1$. However, we can see that an approximate solution
$m^z\approx 1$ and $m^x\approx 0$ is valid in a wider space of parameters.
Indeed, the solution
\begin{eqnarray}
m^x=\frac{1-s}{s\,p\,\lambda},&\,\mathrm{and}\,\,&m^z=\sqrt{1-\paren{\frac{1-s}{s\,p\,\lambda}}^2}
\end{eqnarray}
fulfills  \eqref{eq:mzbetainfty} and \eqref{eq:mxbetainfty} when
$(1-s)/p\,s\,\lambda\to 0$. This is the F phase we obtained before in equation
\eqref{eq:fFca}. The free energy is obtained plugging these values into equation
 \eqref{eq:free}. For the $p\to\infty$ limit,
\begin{equation}\label{eq:freeFinfty}
f_\mathrm{F}(s,\lambda)|_{p\to\infty}=-s\lambda.  \end{equation}

We consider an alternative solution for $0<m^z<1$.  With this aim, we
rewrite  \eqref{eq:mzbetainfty} in the following way \begin{equation}\label{eq:mzsq}
\caja{(m^z)^2-1}\caja{p\,s\,\lambda
  (m^z)^{p-1}}^2+\lazo{m^z\caja{1-s-s(1-\lambda)k(m^x)^{k-1}}}^2=0.  \end{equation}
In the $p\to\infty$ limit, $p(m^z)^{p-1}\to
0$, and 
\begin{eqnarray}\label{eq:solFp}
m^x=\caja{\frac{1-s}{k\,s\,(1-\lambda)}}^\frac{1}{k-1}& m^z=\sqrt{1-\caja{\frac{1-s}{k\,s\,(1-\lambda)}}^\frac{2}{k-1}}
\end{eqnarray}
is an exact solution to  \eqref{eq:mzsq}, and similarly of
 \eqref{eq:mxbetainfty}, as long as $(1-s)/k\,s\,(1-\lambda)\!<\!s\!\le 1$, or
$1/[1+k\,(1-\lambda)]< s< 1$.\footnote{Again,  the negative solution for $m^x$ is also a valid solution in
  the odd $k$ case but has a higher free energy than \eqref{eq:solFp} due to the change of sign in
  the $(m^x)^k$ term in  \eqref{eq:free}.} This is precisely the F' phase
discussed in section \ref{sec:CA}.
Again, we compute the free energy by plugging the solution
\eqref{eq:solFp} in \eqref{eq:free} and taking the $p\to\infty$ limit
\begin{equation}\label{eq:freeFpinfty}
\left. f_{\mathrm{F`}}(s,\lambda)\right|_{p\to\infty}=-\frac{k-1}{k}\caja{\frac{1-s}{k\,s\,(1-\lambda)}}^\frac{1}{k-1}(1-s),
\end{equation}
which is exactly equal to the one obtained for the QP2
phase \eqref{eq:freeQP2}. 

The solution \eqref{eq:solFp} is also a good approximate solution for $p$
finite (but $p>3$) when $(m^z)^p\to 0$. 
The free energy for this phase is
\begin{equation}\label{eq:freeFp}
 f_{\mathrm{F`}}(s,\lambda)\approx-s\,\lambda\caja{1-\paren{\frac{1-s}{s\,k\,(1-\lambda)}}^{\frac{2}{k-1}}}^{p}-\frac{k-1}{k}\caja{\frac{1-s}{k\,s\,(1-\lambda)}}^\frac{1}{k-1}(1-s),\end{equation}
which, for finite $p$, is always smaller than $f_\mathrm{QP2}$. According to
this observation, except for the $p\to\infty$ limit, the F' phase is always stabler than the QP2
phase. 

We have therefore reproduced the results of section \ref{sec:CA} by a
completely different method. The present method is nevertheless better suited
for generalizations to more complicate problems where the target Hamiltonian
$\hat{H}_0$ cannot be expressed in terms of simple total spins.

\section{Ground state of $\hat V_k$ and its overlap with the ground state of $\hat{H}_0$}\label{sec:Vkground}
In this Appendix, we derive the properties of the ground state of $\hat{V}_{k}$ for
$k$ even.
Let us first consider the case with $N$ even.
The ground state of $\hat{H}_0$, $\ket{\phi_0}=\otimes_{i=1}^N\ket{\uparrow}_i^z$,
can be expressed as
\begin{eqnarray}
\ket{\phi_0}&=&\otimes_{i=1}^N\big(\ket{\uparrow}_i^x+\ket{\downarrow}_i^x\big)/\sqrt{2}\nonumber\\
&=&\frac{1}{2^{N/2}}\Big(\ket{\uparrow}_1^x\ket{\uparrow}_2^x\cdots \ket{\uparrow}_N^x
+\ket{\uparrow}_1^x\ket{\uparrow}_2^x\cdots \ket{\uparrow}_{N-1}^x\ket{\downarrow}_N^x\nonumber\\
&&+\cdots +\ket{\downarrow}_1^x\ket{\downarrow}_2^x\cdots \ket{\downarrow}_N^x \Big).
\end{eqnarray}
This last expression has $2^N$ terms, in which the partial sum of terms with a half of the sites
having $\ket{\uparrow}_i^x$ and the other half $\ket{\downarrow}_i^x$ is nothing but the
ground state of $\hat{V}_k$ in the $S=N/2$ sector $\ket{\phi_k}$, up to a normalization,
\begin{eqnarray}
\ket{\phi_k}&=&a \Big(\ket{\uparrow}_1^x\ket{\uparrow}_2^x\cdots\ket{\uparrow}_{N/2}^x
 \ket{\downarrow}_{N/2+1}^x\cdots\ket{\downarrow}_N^x\nonumber\\
&&+\cdots +\ket{\downarrow}_1^x\ket{\downarrow}_2^x\cdots \ket{\downarrow}_{N/2}^x\ket{\uparrow}_{N/2+1}^x
\cdots\ket{\uparrow}_N^x \Big).
\end{eqnarray}
It is easy to check from the number of terms in the above equation
that the normalization condition is $a^2\displaystyle{N \choose N/2}=1$.
We thus have
\be
\left\langle\phi_0|\phi_k\right\rangle=\frac{a}{2^{N/2}}{N\choose N/2}
=\frac{1}{2^{N/2}}\sqrt{{N\choose N/2}}.
\ee
For large $N$, 
\be
\log
|\left\langle\phi_0|\phi_k\right\rangle |^2=\log\caja{2^{-N}\frac{N!}{\paren{\frac{N}{2}!}^2}}\approx
-\frac{1}{2}\log N +\log\sqrt{\frac{2}{\pi}},
\ee
which means that the overlap decreases only polynomically with $N$ as $\sim
N^{-1/2}$.

The case of odd $N$ can be analyzed similarly with $\displaystyle {N\choose N/2}$ replaced
by  $\displaystyle {N\choose (N+1)/2}$ or $\displaystyle {N\choose (N-1)/2}$.
\section*{References}
\bibliography{biblio}
\bibliographystyle{iopart-num}
\end{document}